\documentclass[aps,pra,reprint,nofootinbib]{revtex4-2}

\usepackage{graphicx}
\usepackage{dcolumn}
\usepackage{bm}
\usepackage{cancel}
\usepackage{hyperref}
\usepackage{comment}
\usepackage{mathtools}
\usepackage{empheq}
\usepackage{bbm}
\usepackage{mathrsfs}
\usepackage{amsfonts}
\usepackage{color} 
\usepackage{multirow}
\usepackage{pbox}
\usepackage{xcolor} 
\usepackage{makecell}
\usepackage{braket}
\usepackage{caption}
\usepackage{enumerate}
\usepackage{physics}

\newcommand{\psiF}[2]{\psi_{#1}^{#2}}
\newcommand{\psiDag}[2]{\psi_{#1}^{#2\,\dagger}}
\hypersetup{
    colorlinks=true,
    linkcolor=blue,
    filecolor=magenta,      
    urlcolor=cyan,
}
\usepackage{scalerel,stackengine}
\usepackage{soul}
\soulregister\cite7
\soulregister\eqref7

\newcommand{\add}[1]{\textcolor{teal}{\ul{#1}}}

\usepackage[OT2,T1]{fontenc}
\DeclareSymbolFont{cyrletters}{OT2}{wncyr}{m}{n}
\DeclareMathSymbol{\Sha}{\mathalpha}{cyrletters}{"58}

\usepackage{cleveref}

\makeatletter 

\renewcommand\onecolumngrid{
\do@columngrid{one}{\@ne}%
\def\set@footnotewidth{\onecolumngrid}
\def\footnoterule{\kern-6pt\hrule width 1.5in\kern6pt}%
}

\renewcommand\twocolumngrid{
        \def\footnoterule{
        \dimen@\skip\footins\divide\dimen@\thr@@
        \kern-\dimen@\hrule width.5in\kern\dimen@}
        \do@columngrid{mlt}{\tw@}
}%

\makeatother    
\newcommand{\vp}{{\vec{p}}}

\newcommand{\be}{\begin{equation}}
\newcommand{\ee}{\end{equation}}
\newcommand{\ben}{\begin{equation*}}
\newcommand{\een}{\end{equation*}}

\newcommand{\Zb}{\mathbb{Z}}
\newcommand{\Hc}{\mathcal{H}}
\newcommand{\Oc}{\mathcal{O}}
\newcommand{\PsiB}[1]{\boldsymbol{\Psi}_{#1}}
\newcommand{\barPsiB}[1]{\boldsymbol{\bar{\Psi}}_{#1}}
\newcommand{\Exp}[1]{\text{Exp}\!\left[#1\right]}

\begin{document}

\title{Fermion Doubling in Dirac Quantum Walks}

\author{Chaitanya Gupta}
 \email{chai.gupta@bristol.ac.uk}
\author{Anthony J. Short}
 \email{Tony.Short@bristol.ac.uk}
\affiliation{
 H.H. Wills Physics Laboratory, University of Bristol, Tyndall Avenue, Bristol BS8 1TL, U.K}
\date{\today}

\graphicspath{ {./images/} }

\begin{abstract}
We consider discrete spacetime models known as quantum walks, which can be used to simulate Dirac particles. In particular we  look at fermion doubling in these models, in which high momentum states yield additional low energy solutions which behave like Dirac particles. The presence of doublers carries over to the `second quantised' version of the walks represented by quantum cellular automata, which may lead to spurious solutions when introducing interactions. Moreover, we also consider pseudo-doublers, which have high energy but behave like low energy Dirac particles, and cause potential problems regarding the stability of the vacuum. To address these issues, we propose a family of quantum walks, that are free of these doublers and pseudo-doublers, but still simulate the Dirac equation in the continuum limit. However, there remain a small number of additional low energy solutions which do not directly correspond to Dirac particles. 
While the conventional Dirac walk always has a zero-probability for the walker staying at the same point after a single time step, we obtain the family of walks by allowing this probability to be non-zero.
\end{abstract}

\maketitle

\section{Introduction}
Quantum Cellular Automata (QCA) and Quantum Walks (QW) have been shown to provide a discrete spacetime model of quantum field theories, with the latter being a single particle limit of the former \cite{farrelly2020review,farrelly2014causal,brun2025quantum}. While these models provide an interesting avenue for quantum simulations of particle physics, they may also provide insight into physical models of reality itself \textemdash if reality is indeed discrete. 

Of particular interest are discrete models that are able to simulate free Dirac particles. In particular, the Dirac QW is able to simulate free Dirac particles in the continuum limit \cite{succi1993lattice,Bialynicki-Birula,TonyWalk}. Its `second quantised' cousin, the Dirac QCA simulates free Dirac particles at a field theoretical level \cite{d2017quantum,brun2020quantum,farrelly2014causal}. Recent papers have introduced interactions to the Dirac QCA to simulate other theories such as QED \cite{arrighi2020quantum,eon2023relativistic}. Extensions to introduce background electromagnetic and gravitational fields have also been developed \cite{bggravity,Montero16Background,arrighi2016quantumGravity,Arrighi2018quantumwalkingin}. Moreover, renormalisation of QCA models has also been studied, with the potential application of simulating more phenomena and reducing the complexity of existing models \cite{trezzini2025renormalisation,fermionicrenormalisation}. There has also been an increasing interest in experimental implementation of these models \cite{qiang2024quantum,Zhou_2021Experiment,wang2013physical,DiracExperimentHuertaAlderete2020}.

Now, similarly to other discrete theories  \add{\cite{rothe2012lattice,wilson1974confinement,kogut1975hamiltonian,susskind1976strong,montvay1994LFT}}, if we were to attempt scattering calculations using these models, we would be likely to run into trouble if there was any fermion doubling. By fermion doubling, we mean the presence of high momentum particles that have low energy, and behave like additional species of Dirac fermions in the continuum limit. 

If we look at the Dirac QW in 1+1-D, we do not get fermion doubling \cite{bakircioglu2025fermion}, but nevertheless, find high momentum particles that behave like additional species of Dirac particles in the continuum limit but do not have low energy. We call these particles pseudo-doublers as they do not have the same energy as the low-momentum particles, hence, are not doublers in a traditional sense. Moreover, in the 3+1-D walk, in addition to the presence of these pseudo-doublers, we also have fermion doubling in a more traditional sense. We argued in Ref.~\cite{Gupta2025diracvacuumin} that  pseudo-doublers are likely to cause problems in interacting models by making the vacuum state highly unstable.

In this work, we present a new family of quantum walks for 1+1D and 3+1D, which yield the Dirac equation in the continuum limit, with the aim of eliminating fermion doubling and pseudo-doubling. 
We do so by constructing a very general unitary which moves a particle by at most one lattice site in each direction in a single time step. Now, in the conventional walk, the probability of the particle to stand still is always zero. We relax this condition to obtain a broader category of walks. By choosing appropriate values of a parameter,  we find that, we are able to get rid of all fermion doubling and pseudo-doubling. However, in 3+1-D, the walk still has some extraneous low-energy solutions which do not act as traditional fermion doublers.

There has been some previous interest in the literature in exploring doubling in Dirac quantum walks. Ref. \cite{jolly2023twisted} explored an interesting family of twisted quantum walks in 1+1-D, in which a particle moves multiple lattice sites in a single time step, that yields both the Dirac equation and a Laplacian operator in the continuum walk. These walks admitted fermion doubling, in a manner that is very similar to lattice field theories, and the work suggested an approach to solve the doubling by modifying the walks. Moreover, Ref. \cite{bakircioglu2025fermion} explored the doubling in a Dirac walk, and suggested interpreting the doublers as different flavours. While, it would be interesting to explore such an approach, it is still not clear how such an identification would lead to correct interactions in the continuum theory.
 
The paper is organised as follows. In Section \ref{sec:normalwalk}, we introduce the conventional Dirac QWs in 1+1-D and 3+1-D and discuss the presence of doublers and pseudo-doublers. In Section \ref{sec:slowwalk}, we introduce the parametrised family of Dirac QWs in 1+1-D and 3+1-D which reduce the doublers and pseudo-doublers. We have provided the details of the QCA based on these walks in the Appendix. Note that, throughout the paper, we shall set $\hbar = 1$, and let $c$ be the speed of light.

\section{Fermion Doubling in Conventional Dirac Walks}
\label{sec:normalwalk}
\subsection{1+1-D Conventional Dirac Walk}
Consider a particle on a 1-D lattice with lattice spacing $\delta x$. The position of the particle is described by the Hilbert space $\Hc_{\Zb}$ which is spanned by basis $\{\ket{n}|n\in\Zb\}$. Suppose the particle has a two-dimensional internal state with the corresponding Hilbert space $\Hc_{\text{S}}\cong\mathbb{C}^2$.  

Let $\gamma_{\mu}$ with $\mu\in\{+,-,0\}$ be operators on $\Hc_{\text{S}}$. Suppose we wish to define a quantum walk of the form 
\be
    T = \gamma_{+} S + \gamma_{0} + \gamma_{-} S^\dagger
    \label{eq:unitaryabstract}
\ee
with $T$ being a unitary operator on $\Hc_{\text{S}}\otimes\Hc_{\Zb}$ describing the evolution of the particle\footnote{Note that to avoid cumbersome notation, when considering operators on a subsystem (such as $\gamma_+$), we use the same symbol to represent the action of the operator on the appropriate subspace and on the whole Hilbert space. In the latter case, there is an implicit tensor product with the identity operator on the other subsystems.} in a single time step $\delta t$.
Here, $S$ is the shift operator which acts on the position states as $S\ket{n}=\ket{n+1}$ for every $n\in\Zb$. 

It is clear that $T$ is local ($\bra{m+n}T\ket{n}= 0$ if $m \notin \Lambda$ where $\Lambda \subset \Zb$ is finite). Thus, for $T$ to define a valid quantum walk, it needs to be unitary. We have to pick values of $\gamma_{k}$ such that it is so. 

The most common choice in literature is to take $\gamma_{0} = 0$, while letting $\gamma_{+}$ and $\gamma_{-}$ be projection operators onto orthogonal spaces. This choice gives us the conventional Weyl walk which is a well-known walk 
\cite{succi1993lattice,Bialynicki-Birula}. 

To make this choice explicitly, let $\{\ket{\uparrow_z},\ket{\downarrow_z}\}$ be some orthonormal basis for $\Hc_{\text{S}}$. Then, we let $\gamma_{+}=\ketbra{\uparrow_z}$, $\gamma_{-}=\ketbra{\downarrow_z}$ and $\gamma_0=0$. Substituting this choice in Eq.~\eqref{eq:unitaryabstract}, we get
\be
    T = \ketbra{\uparrow_z}S + \ketbra{\downarrow_z}S^\dagger
    \label{eq:originalDirac}
\ee
which we can check is indeed unitary ($T^\dagger T = TT^\dagger=I$). 

We can extend the model to massive particles by introducing 
\be
    W = \Exp{-iM\delta t}
    \label{eq:massUnitary}
\ee
where $M$ is Hermitian operator on $\Hc_{\text{S}}$ with the largest absolute value of its eigenvalues being $\alpha$.
Now, we let our walk be described by
\be
    U = WT.
    \label{eq:TotalUnitary}
\ee
Here, $U$ is indeed unitary and local, and thus, a valid quantum walk \cite{farrelly2020review}. It also has some additional symmetries: translational invariance ($[S,U] = 0$) and unitary equivalence to its Hermitian adjoint ($\sigma_y U \sigma_y^\dagger = U^\dagger$).

To consider the continuum limit of this walk, we define momentum eigenstates and the momentum operator as
\be
\!\ket{p} = \sqrt{\frac{\delta x}{2\pi}}\sum_n e^{i p n\, \delta x}\ket{n} \,\text{and}\,
P =\! \int_{-\pi/\delta x}^{\pi/\delta x} \!\!dp \, p\ketbra{p},
\ee
where we have chosen the momentum to run from $-\pi/\delta x$ to $\pi/\delta x$.

Now, using the fact that $S= e^{-iP\,\delta x}$, we can rewrite $T$ from Eq.~\eqref{eq:unitaryabstract} as
\begin{align}
    T &= 
    \ketbra{\uparrow_z}e^{-iP\,\delta x}+ \ketbra{\downarrow_z}e^{iP\,\delta x}
    \nonumber\\
    &= \Exp{-i P \sigma_z \delta x}
    \label{eq:conventionalmomentum}.
\end{align}
where $\sigma_z \equiv \ketbra{\uparrow_z}-\ketbra{\downarrow_z}$.

Now, let us look at the continuum limit. Consider a state $\ket{\psi_p} = \ket{p}\otimes\ket{s}$. Suppose we have $p\delta x \ll 1$ and $\alpha \delta t \ll 1$, then
\begin{align}
    U \ket{\psi_p} &= WT  \ket{\psi_p}  
    \nonumber\\
    &= \Exp{-iM\delta t}\Exp{-iP\sigma_z\delta t} \ket{p}\ket{s}
    \nonumber\\
    &=\Exp{-iM\delta t}\Exp{-ip\sigma_z\delta t} \ket{p}\ket{s}
    \nonumber\\
    &=(1-iM\delta t + \Oc(\delta t^2))(1-ip\sigma_z\delta x + \Oc(\delta x^2))\ket{p}\ket{s} 
    \nonumber\\
    &=(1-iM\delta t-ip\sigma_z\delta x + \Oc(\delta x^2,\delta x\delta t,\delta t^2))\ket{p}\ket{s}
    \nonumber\\
    &=(1-iM\delta t-iP\sigma_z\delta x + \Oc(\delta x^2,\delta x\delta t,\delta t^2))\ket{\psi_p}
    \label{eq:powerexpandconventional}.
\end{align}

Looking at the expansion, we can write an effective Hamiltonian as, 
\be
    H_{\text{eff}} = M+Pv\sigma_z.
\ee
in the continuum limit where $v \equiv \delta{x}/\delta{t}$, such that 
\be
U \ket{\psi_p} = e^{-i H_{\text{eff}} \delta t} \ket{\psi_p}  +  \Oc(\delta x^2,\delta x\delta t,\delta t^2). \label{eq:onetimestepU}
\ee

Now, with $\sigma_x\equiv\ketbra{\uparrow_z}{\downarrow_z}+\ketbra{\downarrow_z}{\uparrow_z}$, we choose $M=mc^2\sigma_x$ and $v=c$ where $m$ is the mass of the particle. This gives us
\be
    H_{\text{eff}}= mc^2\sigma_x+Pc\sigma_z.
\ee
which is the 1+1-D Dirac Hamiltonian. Moreover, with the above choice of $M$, $U$ describes the conventional Dirac QW. We wish to emphasise than when the mass and momentum of the particle are small ($mc^2 \delta t \ll 1$ and $p\,\delta x \ll 1$), the conventional Dirac quantum walk can be described by the Dirac Hamiltonian; thus, the walk simulates free Dirac fermions in the continuum limit. Note that \eqref{eq:onetimestepU} shows that the deviation   from the continuum model in the unitary for one time-step is second order in small quantities. Hence, even over a  significant time interval (comprised of $\Oc\left(\frac{1}{\delta t}\right)$ time steps), the overall deviation between the discrete and continuous model will be $\Oc\left(\delta t\right)$ and thus small as long as $\delta t$ is sufficiently small.

While we have defined a notion of an effective Hamiltonian for only small momentum and mass, we can define the energy of the particle in general. Consider the operator
\begin{align}
    U(p)&\equiv\Exp{-ip\sigma_z\delta x}\Exp{-iM\delta t}
\end{align}
acting on $\Hc_{\text{S}}$. Let $\ket{s_p}$ be eigenstates of $U(p)$ with the respective eigenvalues being $e^{-iE_{p}\delta t}$. Restricting $E_p$ in $(-\pi/\delta t, \pi/\delta t]$, and thereby choosing a branch cut, we can identify $E_{p}$ as the possible energy eigenvalues of a particle in the energy eigenstate $\ket{p}\ket{s_p}$. Moreover, because of the unitary equivalence of $U$ to $U^\dagger$, for every energy eigenvalue $E_{p}>0$, $-E_{p}<0$ is also a possible energy eigenvalue of the walker.

Now, having defined a notion of energy for all momentum, and having seen that the walk behaves like a Dirac particle for small momentum (for a choice of $v$ and $M$), we wish to know if the walk behaves like Dirac particles at any other point in momentum space.

Let us consider $p=\pm\pi/\delta x+\eta$. We have,
\be
    U(\pm\pi/\delta x+\eta) = -U(\eta).
\ee

As this differs from $U(\eta)$ only by a global phase, particles with momentum in a small ball around $\pm\pi/\delta x$ with $\alpha\delta x\ll 1$ also behave like low-momentum particles, as was pointed out in \cite{Gupta2025diracvacuumin,bakircioglu2025fermion}. However, they have energy near $E=\pm\pi/\delta t$. As their energy is different from low energy particles, we do not exactly have fermion doubling. We call these particles pseudo-doublers.

Going back to our model, we note that, for the appropriate choice of $\delta t$ and $M$, particles near both $p=0$ ($E=0$) and $p=\pm\pi/\delta x$ ($E=\pm\pi/\delta t$) behave like Dirac particles. If we were to second quantise this model to a QCA, the presence of these pseudo-doublers carries over (See Appendix \ref{app:QCAFree} for details of the QCA). In \cite{Gupta2025diracvacuumin}, we had argued that, the presence of such solutions causes the Dirac vacuum to be unstable. 

\subsection{3+1-D Conventional Dirac Walk}
\label{sec:3dnormal}

Let us now look at a higher dimensional case. We shall start by constructing the Weyl walk, and then introduce a mass term to yield the Dirac walk. Consider a 3-D cubic lattice with the lattice spacing being $\delta x$. A particle lives on the lattice and is described by the Hilbert space $\Hc_{\text{S}}\otimes\Hc_{\Zb^3}$ with $\Hc_{\Zb^3} = \text{span}\{\ket{\vec{n}}|\vec{n}\in \Zb^3\}$, and $\Hc_{\text{S}}\cong\mathbb{C}^2$ as before. 

We wish to define a quantum walk describing the evolution of the particle. To that goal, let $\gamma^j_{\mu}$ be linear operators on $\Hc_{\text{S}}$ with $\mu\in\{+,0,-\}$ and $j\in\{x,y,z\}$. Then, we let
\be
    K = K_zK_yK_x
    \label{eq:unitaryabstract3d}
\ee
define a walk which describes the evolution of the system in time step $\delta t$, and where
\begin{align}
    K_j =  \gamma^j_{+} S_j + \gamma^j_{0}+ \gamma^j_{-} S_j^\dagger
    \label{eq:KjAbstract}
\end{align}
are operators on $\Hc_{\text{S}}\otimes\Hc_{\Zb^3}$. Here, $S_j$ is a shift operator in the $j$-direction; for instance, $S_x\ket{(n_x,n_y,n_z)}=\ket{(n_x+1,n_y,n_z)}$. 

Again, $K$ is local and translationally invariant. We now wish to pick our $\gamma_{\mu}^j$ to ensure unitarity of $K$. This can be done by ensuring that each $K_j$ is unitary. A common choice in literature is to pick \cite{farrelly2014causal}
\begin{align}
    \gamma_{+}^{j} = \ketbra{\uparrow_j}, \gamma_{-}^{j} = \ketbra{\downarrow_j}, \text{ and } \gamma_{0}^{j} = 0
\end{align}
where $\ket{\uparrow_j}$ and $\ket{\downarrow_j}$ are defined via $\sigma_j\ket{\uparrow_j}=\ket{\uparrow_j}$ and $\sigma_j\ket{\downarrow_j}=-\ket{\downarrow_j}$. Here, $\sigma_y \equiv -i\ket{\uparrow_z}\bra{\downarrow_z}+i\ket{\downarrow_z}\bra{\uparrow_z}$.
With this choice, $K$ becomes the Weyl walk in 3+1-D. 

Let us now go to momentum space. We define the momentum eigenstates and the momentum operators as 
\begin{align}
\ket{\vec{p}} &= \left(\frac{\delta x}{2\pi}\right)^{\tfrac{3}{2}}\sum_{\vec{n}} \Exp{i\vp\cdot\vec{n}\,\delta x}\ket{\vec{n}} \;\text{and}\;
\nonumber\\
\vec{P} &= \int_{\Theta} d^3 p \,\vec{p} \ketbra{\vec{p}},
\end{align}
where
\be
\Theta= (-\pi/\delta x,\pi/\delta x]^3
\ee
and $\vec{P}$ is understood to be a vector of operators.

We can now write the shift operators as $S_j = \Exp{-iP_j\delta r_j}$. This allows us to rewrite $K_j$ as
\begin{align}
    K_j&=\Exp{-i\vec{P}\cdot\vec{\sigma}\delta x}.
\end{align}
where $\vec{\sigma}=(\sigma_x,\sigma_y,\sigma_z)^\mathbf{T}$.

Then, we have,
\be
    K = \Exp{-iP_z\sigma_z\delta x}\Exp{-iP_y\sigma_y\delta x}\Exp{-iP_z\sigma_z\delta x}.
\ee

\begin{figure}
	\centering
        \centering
 \includegraphics[width=0.95\linewidth]{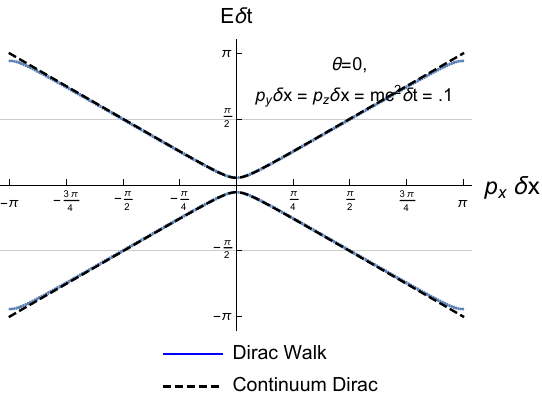}
 \caption{Plot of the dispersion relation for the 3+1-D conventional Dirac quantum walk and for the continuum theory.}
 \label{fig:3DGraphNormal}
\end{figure}

Let us now look at the continuum limit. Similar to the 1+1-D case, we consider $\ket{\psi_\vp}=\ket{\vec{p}}\ket{s}$ where $\vp\in\Theta$ and $\ket{s}\in\Hc_{\text{S}}$. Then, assuming that $p_j\delta x\ll 1$, we have

\begin{align}
K_zK_yK_z \ket{\psi_\vp}&=
\prod_{j}\Exp{-i P_j\sigma_j\delta x}\ket{\vp\,}\ket{s}
\nonumber
\\
&=
\prod_{j}(I-ip_j\sigma_j\delta x+\Oc(\delta x^2))\ket{\vp\,}\ket{s}
\nonumber
\\
&=
\left( I-i\sum_j p_j\sigma_j\delta x +\Oc(\delta x^2)\right)\ket{\vp\,}\ket{s}
\nonumber
\\
&=
\left( I-i\vec{P}\cdot\vec{\sigma}\,\delta x +\Oc(\delta x^2)\right)\ket{\psi_\vp}
\end{align}
where $\prod_j$ is understood to be an ordered product.

Again, this allows to define an effective Hamiltonian for this regime,
\be
    H^{\text{eff}} = \vec{P}\cdot\vec{\sigma} v
\ee
where $v = \delta x/\delta t$ as before. Now, if we let $v=c$, we get the Weyl equation.

We can again define energy $E_{\vp}\in(-\pi/\delta t,\pi/\delta t]$ as before by letting $\Exp{-iE_{\vp}\,\delta t}$ be the eigenvalues of $K(\vp\,)$ where
\be
    K(\vp\,) = \Exp{-ip_z\sigma_z\delta x}\Exp{-ip_y\sigma_y\delta x}\Exp{-ip_x\sigma_x\delta x}.
\ee

Having looked at its low-momentum limit, we wish to know if for any other $p$ does the model behave like its low-momentum limit. In fact, for $\vp=(\pm\pi/\delta x,\pm\pi/\delta x,0)$ and its respective permutations, we have
\be
    K(\vp+\vec{\eta}\,) = K(\vec{\eta}\,).
\ee
Thus, particles near the point above also behave like low-momentum particles and have energy near $0$. This is an example of fermion doubling.

Also, for $\vp=(\pm\pi/\delta x,\pm\pi/\delta x,\pm\pi/\delta x)$ and  $\vp=(\pm\pi/\delta x,0,0)$ with its respective permutations, we have
\be
    K(\vp+\vec{\eta}\,) = - K(\vec{\eta}\,).
\ee
Thus, particles near this point also behave like low-momentum particles but have energy near $\pm\pi/\delta t$. This is an example of pseudo-fermion doubling. Note that these doublers and pseudo-doublers have been discussed to some extent in \cite{bakircioglu2025fermion}.

We also have another set of points $\vp=(\pm\pi/2\delta x,\pm\pi/2\delta x,\pm\pi/2\delta x)$ which behave similarly to Weyl particles of opposite chirality (i.e. with  $H^{\text{eff}} = -\vec{P}\cdot\vec{\sigma} c$), with energy close to either $0$ or $\pi/\delta t$ depending on the signs of the momenta \cite{Gupta2025diracvacuumin}. These are investigated in more detail in Appendix \ref{app:weirdpoint}. 

We now want to introduce a mass term. To do so, we need to double the internal space. Let the particle's state be now described by $\Hc^\prime=\Hc_{\Zb^3}\otimes\Hc_{\text{int}}$ where $\Hc_{\text{int}}=\Hc_{\text{S}}^{\otimes 2}\cong\mathbb{C}^4$. 

Then, introduce the unitary,
\be
W = \Exp{-i m c^2 \beta\, \delta t}
    \label{eq:mass3d}
\ee
where 
\be
    \beta = \begin{pmatrix}
        0 & \mathbb{I} \\
        \mathbb{I} & 0
    \end{pmatrix}
\ee
and $m$ is the mass of the particle.
Then, the walk,
\be
    U = W  \begin{pmatrix}
        K_zK_yK_x & 0 \\
        0 & K_z^\dagger K_y^\dagger K_x^\dagger
    \end{pmatrix}
    \label{eq:3dwalkdefine}
\ee
describes the 3+1-D Dirac walk \cite{TonyWalk,LeonardWalk}. 

In the continuum limit (which can be obtained by considering a state with definite momentum $\vp$ such that $p_j\delta x\ll 1$ and letting $mc^2\delta t\ll 1$), we have

\begin{align}
 H_{\text{eff}} = 
 \begin{pmatrix}
     \vec{P}\cdot\vec{\sigma}v & \!\!\!mc^2 \\
    mc^2 & \!\!\!-\vec{P}\cdot\vec{\sigma}v
 \end{pmatrix}\!.
 \label{eq:abstractHamil3+1d}
\end{align}
Again, setting $v=c$, we get the 3+1-D Dirac Hamiltonian. Thus, for small momentum and mass ($p_j\delta x \ll 1$ and $mc^2\delta t \ll 1$), the above walk describes a Dirac fermion.

Similar to previous cases, we can write $U(p)$ as an operator on $\Hc_{\text{int}}$ to describe the walk in momentum space. Then, we can define the energy eigenvalues associated with the walk as before (Figure \ref{fig:3DGraphNormal}).

We have the same doublers and pseudo-doublers as in the case of the Weyl walk except for the one at $\vp=(\pm\pi/2\delta x,\pm\pi/2\delta x,\pm\pi/2\delta x)$. Now, at this point, the Dirac walk does not have a good continuum limit. While we do not consider it as a doubler or a pseudo-doubler, its energy eigenvalues do lie near both $E=0$ and $E=\pm\pi/\delta t$ (supposing small $mc^2\delta t)$.

Having presented these walks, we note that we can second quantise these walks to have a QCA model (Appendix \ref{app:QCAFree}). It is highlighted in \cite{Gupta2025diracvacuumin} that the pseudo-doublers cause problems with the stability of the Dirac vacuum in the QCA model, seemingly allowing processes in which a particle and antiparticle are created while releasing energy. We also expect the normal doublers to interfere with any scattering calculations. We shall now construct a new walk to get rid of both the doublers and pseudo-doublers.

\section{A Family of Dirac Walks}
\label{sec:slowwalk}

In the previous section, we had constructed the walks from Eq.~\eqref{eq:unitaryabstract} and Eq.~\eqref{eq:unitaryabstract3d} such that the amplitude for the particle to stay in the same place is zero. However, we can pick $\{\gamma_j\}$ and $\{\gamma^\mu_j\}$ differently to allow the particle to stay where it is with some non-zero probability. Let us first look at the 1+1-D case.

\subsection{1+1-D}

\begin{table*}
\setlength{\tabcolsep}{10pt} 
\renewcommand{\arraystretch}{1.5} 
\begin{tabular}{ | r | c | c | } 
  \hline
   & \bf{Conventional 1+1-D Dirac Walk} & \bf{Family of New 1+1-D Dirac Walks}  \\ 
  \hline
  Parameters & none\footnote{We consider $m$ to be fixed, and the lattice spacing $\delta x$ to be an artefact of the lattice rather than the walk.} &  $\theta$ \\ 
  \hline
  Symmetries & locality, homogeneity, unitarity & locality, homogeneity, unitarity \\
  \hline
  \rule{0em}{1.5em}\makecell[r]{Relationship between \\ lattice spacing and time step} &$\delta x = c\delta t$ & $\cos(\theta)\delta x = c \delta t$\\ 
  \hline
  Continuum Limit Behaviour & Dirac Hamiltonian & Dirac Hamiltonian \\ 
  \hline
  Doublers & none & none\\
  \hline
  Pseudo-doublers & at $p\,\delta x = \pm\pi$ &  none for $\theta \neq 0$
  \\ 
  \hline
   Range of Eigenvalues & $\abs{E\delta t} \leq \pi$ & $\abs{E\delta t} \leq \pi - 2\theta + mc^2\delta t$
   \\
   \hline
\end{tabular}
    \caption{Comparison between the conventional and the new Dirac walks in 1+1-D. The new walk is rather a family of walks parametrised by $\theta \in (-\pi/2,\pi/2)$. When $\theta=0$, the new walk reduces to the conventional walk. In the conventional walk, during each time step, the walker either moves left or right. In contrast, for the new family of walks, the walker may also stay still along with moving right or left in a single time step. Crucially, for $\theta\neq 0$, the new walk doesn't have any doublers or pseudo-doublers. Moreover, introducing $\theta$ allows us to tune the parameter to restrict the energy eigenvalues.}
    \label{tbl:1d}
\end{table*}

\begin{figure}
	\centering
		\includegraphics[width=\linewidth]{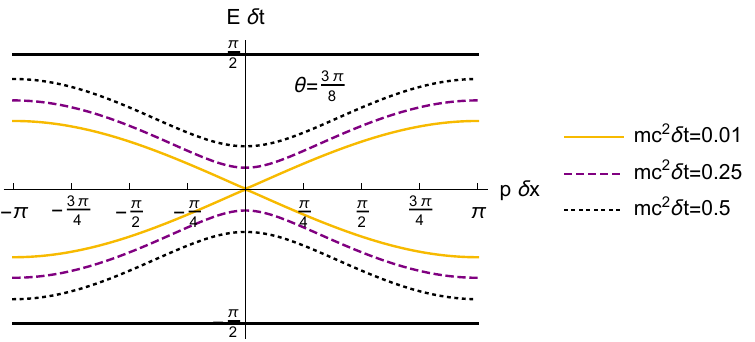}
 \caption{Plot of the dispersion relation of the parametrised 1+1-D Dirac walk with $\theta = 3\pi/4$ and different values of $m$. With this choice of $\theta$, we can see that the energy is always less than $\pi/(2\delta t)$ for the plotted values of mass ($m$).}
 \label{fig:1dplots}
\end{figure}

\begin{figure}
	\centering
	\includegraphics[width=\linewidth]{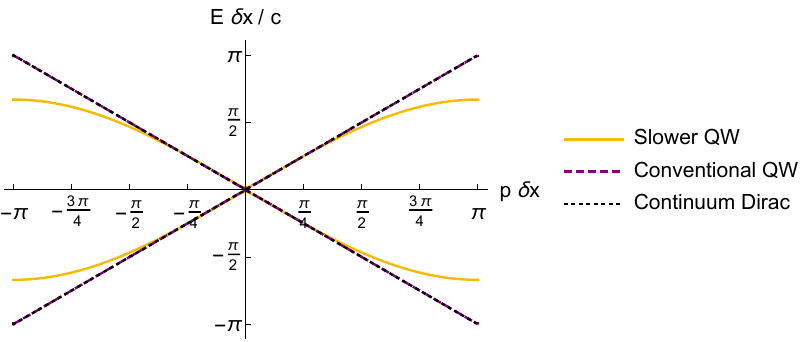}
 \caption{We compare the dispersion relation of the parametrised QW ($\theta = 3\pi/4$) with the conventional walk ($\theta = 0$) and the continuum case (QFT) for $mc^2\delta t=0.02$. It is evident that the plots agree for small $p\,\delta x$.}
 \label{fig:1dplotsCompare}
\end{figure}

\label{sec:slowwalk1d}
Again, we start with the following unitary,

\be
    T = \gamma_{+} S + \gamma_{0} + \gamma_{-} S^\dagger
    \tag{\ref{eq:unitaryabstract}}.
\ee

Earlier, we had set $\gamma_{0}=0$ (thereby ensuring that the probability for the particle to  stand still, after a single time step, is zero) and chosen $\gamma_{+}$, $\gamma_{-}$ to be orthogonal projection operators to get the conventional Dirac walk. Now, we shall pick a choice of $\{\gamma_j\}$ to get the desired continuum limit while maintaining unitarity (and which we will show allows us to remove any pseudo-doublers).

To this goal, consider subspaces $A,B \subset \Hc_{\text{S}}$. Now, let $\Pi_{a}$ and $\Pi_{b}$ be projector operators into the subspaces $A$ and $B$ respectively. Also, then, $\Pi_{\bar{a}}\equiv I - \Pi_a$ is a projector into the complimentary subspace $A^c$. Similarly, we can define  $\Pi_{\bar{b}}$. Let us now make the choice
\begin{align}
    \gamma_{+} &= \Pi_{a}\Pi_{b}
    \nonumber\\
    \gamma_{-} &= \Pi_{\bar{a}}\Pi_{\bar{b}}
    \nonumber\\
    \gamma_{0} & = 1 - \gamma_{+} - \gamma_{-} = (\Pi_{a}\Pi_{\bar{b}}+\Pi_{\bar{a}}\Pi_{{b}}).
    \label{eq:gammaValue}
\end{align}
With the above choice of $\{\gamma_{j}\}$, we can  check that indeed $T^\dagger T = TT^\dagger = I$ (Appendix \ref{app:unitarity}). 

Now, using the fact that $S= e^{-iP\,\delta x}$ and Eq.~\eqref{eq:gammaValue}, we can rewrite $T$ as 
\begin{align}
    T &= 
    \gamma_{+} e^{-iP\,\delta x} + \gamma_{0}+ \gamma_{-}e^{iP\,\delta x}
    \nonumber\\
    &=\Pi_{a}\Pi_{b} e^{-iP\,\delta x} + (\Pi_{a}\Pi_{\bar{b}}+\Pi_{\bar{a}}\Pi_{{b}})+ \Pi_{\bar{a}}\Pi_{\bar{b}} e^{iP\,\delta x}
\label{eq:unitaryabstract2}.
\end{align}

Again, we introduce a mass term via $W$ from Eq.~\eqref{eq:massUnitary}, and let our walk be given by $U=WT$.

Then, after taking the continuum limit, we can read off $H_{\text{eff}}$ as 
\be
    H_{\text{eff}} \equiv (\gamma_{+}-\gamma{-}) P v + M =  (\Pi_a\Pi_b-\Pi_{\bar{a}}\Pi_{\bar{b}}) P v + M
    \label{eq:hamilAbstract}
\ee
with $v=\delta x/\delta t$ as before. 

Consider the operator
\begin{align}
U(p)&\equiv\left(\gamma_{+} e^{-ip\,\delta x} + \gamma_{0}+ \gamma_{-} e^{ip\,\delta x}\right)
\nonumber\\
&\times\Exp{-iM\delta t}
\label{eq:Up1d}
\end{align}
acting on the spin degrees of freedom. Again, if let $e^{-iE_{p}\delta t}$ be the eigenvalues of $U(p)$, we can identify $E_{p}$ as energy while restricting $E_{p}$ to $(-\pi/\delta t,\pi/\delta t]$.

 To simplify the situation, whilst still allowing some tunability of the walk, we now specialise to particular choices for $A$ and $B$ parametrised by the  variable $\theta$. Let $-\pi/2< \theta <\pi/2$ and introduce the unitary, $R_{\theta}=e^{-i\theta\sigma_x/2}$. Then, consider the following operators,
\be
    \sigma_\theta \equiv R_{\theta}\sigma_z R_{\theta}^\dagger = \cos\theta\,\sigma_z-\sin\theta\,\sigma_y
\ee
where  $\sigma_y \equiv -i\ketbra{\uparrow_z}{\downarrow_z}+ i\ketbra{\downarrow_z}{\uparrow_z} $.

We define $\ket{\uparrow_\theta}$ and $\ket{\downarrow_\theta}$ as 
\begin{align}
    \ket{\uparrow_\theta} &=  R_{\theta}\ket{\uparrow_z} =\cos(\theta/2)\ket{\uparrow_z}-i\sin(\theta/2)\ket{\downarrow_z},
    \nonumber\\
    \ket{\downarrow_\theta} &= R_{\theta}\ket{\downarrow_z}=-i\sin(\theta/2)\ket{\uparrow_z}+\cos(\theta/2)\ket{\downarrow_z}.
\end{align}

such that they are the eigenstates of $\sigma_\theta$, i.e.,
\be
    \sigma_\theta\ket{\uparrow_\theta} =\ket{\uparrow_\theta}
    \text{,  and  }
    \sigma_\theta\ket{\downarrow_\theta} =-\ket{\downarrow_\theta}.
\ee

Then, we take 
\be
    \Pi_a = \ketbra{\uparrow_\theta}{\uparrow_\theta}
    \text{, and }
    \Pi_b = \ketbra{\uparrow_{-\theta}}{\uparrow_{-\theta}}.
\ee

Then, it follows that
\be
\Pi_a\Pi_b-\Pi_{\bar{a}}\Pi_{\bar{b}} = \cos(\theta)\sigma_z.
\ee

Then, from Eq.~\eqref{eq:hamilAbstract}, it follows that
 \be
    H^{\text{eff}} = Pv\cos(\theta)\,\sigma_z+M
    \label{eq:DiracHamilTheta}
\ee
in the continuum limit. 

Finally, for this walk, we let $c=\cos(\theta)v$. Then, with $M=mc^2\sigma_x$, the effective Hamiltonian for this walk (Eq.~\eqref{eq:DiracHamilTheta}) becomes
the Dirac Hamiltonian. We have plotted the dispersion relation for this walk with the above choices in Figures \ref{fig:1dplots} and \ref{fig:1dplotsCompare}.

We have obtained a family of Dirac QWs in 1+1-D that are parametrised by $\theta$. We can get our original walk (Eq.~\eqref{eq:originalDirac} and Eq.~\eqref{eq:TotalUnitary}) with $\theta=0$. For $\theta\neq 0$, there are no fermion doublers or pseudo-doublers (see Appendix \ref{app:continuum}). We have summarised the differences between our new family of Dirac QWs and the conventional walk in TABLE \ref{tbl:1d}. Moreover, we present  simulations  comparing the behaviour for an initially localised state in the new walk and the conventional walk in Appendix \ref{app:simulation}.

Finally, we wish to remark on the continuum limit of this walk for arbitrary one-dimensional $A$ and $B$. Consider the operator
\begin{align}
    \Gamma \equiv \gamma_{+}-\gamma{-} =\Pi_a\Pi_b-\Pi_{\bar{a}}\Pi_{\bar{b}}
    \nonumber\\
    = \Pi_a + \Pi_b - I
\end{align}
from Eq.~\eqref{eq:hamilAbstract}. 

We can check that $\Gamma$ is traceless and Hermitian. This allows us to let $\pm a$ be the eigenvalues of $\Gamma$ with $a\geq 0$. Moreover, noting that $2 a^2 = \tr{\Gamma^2} \leq 2$ gives us $a\leq 1$. Thus, via an appropriate choice of axis in the Bloch sphere representing $\Hc_{\text{S}}$, we can always write $\Gamma$ to be proportional to $\sigma_z$ with the proportionality constant being a positive number less than (or equal to) one. Thus, in the continuum limit, when $M=0$, we always get the Weyl Hamiltonian. Also, to get the Dirac Hamiltonian in the continuum limit, we can always choose $M$ to be proportional to a Pauli operator in a direction orthogonal to $z$ on the Bloch sphere. 

Thus, for arbitrary one-dimensional $A$ and $B$, we can see that this walk gives the Weyl and Dirac equation for an appropriate choice of $M$. This is not surprising as it was shown by Ref.~\cite{TonyWalk}, that a very generally walk that obeys certain symmetries, gives the Weyl Hamiltonian in the continuum limit (and then adding an additional unitary with the mass term yields the Dirac Hamiltonian). However, to make it easier to study this generalised walk, we specialise our choice of $A$ and $B$ to a one-parameter family. Consequently, we obtain a family of walks, that gives the correct continuum limit, while allowing its behaviour to be tuned using $\theta$ to get rid of fermion pseudo-doubling.

\begin{figure}
    \centering
    \includegraphics[width=\linewidth]{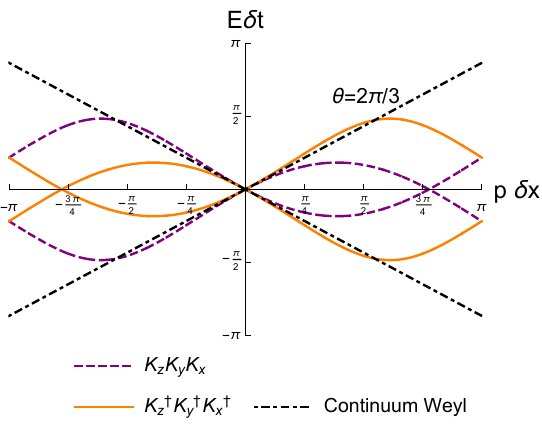}
    \caption{Plot of the dispersion relation of the parametrised 3-D Weyl walks with $\theta= 2 \pi/3$ and $p_x=p_y=p_z=p$. We can see that both the Weyl walks have the doubler at different points. The plot also shows the dispersion relation from the continuum Weyl equation and we can see that the continuum theory agrees with the walks for small $p\delta x$.}
    \label{fig:samep}
\end{figure}

\begin{table*}
\setlength{\tabcolsep}{10pt} 
\renewcommand{\arraystretch}{1.5} 
\begin{tabular}{ | r | c | c | } 
  \hline
   & \bf{Conventional 3+1-D Dirac Walk} & \bf{Family of New 3+1-D Dirac Walks}  \\ 
  \hline
  Parameters & none\footnote{We consider $m$ to be fixed, and the lattice spacing $\delta x$ to be an artefact of the lattice rather than the walk.} &  $\theta$ \\ 
  \hline
  Symmetries & locality, translational invariance, unitarity & locality, translational invariance, unitarity \\
  \hline
  \rule{0em}{1.5em}\makecell[r]{Relationship between \\ lattice spacing and time step} &$\delta x = c\delta t$ & $\cos(\theta)\delta x = c \delta t$\\ 
  \hline
  Continuum Limit Behaviour & Dirac Hamiltonian & Dirac Hamiltonian \\ 
  \hline
  Doublers & at $\vp\,\delta x=(\pm\pi,\pm\pi,0)$ and permutations & none when $\theta \neq 0$\\
  \hline
  \rule{0em}{1.5em}Pseudo-doublers &  \makecell{ at $\vp\,\delta x=(\pm\pi,0,0)$ and permutations, \\ $\vp\,\delta x=(\pm\pi,\pm\pi,\pm\pi)$}  &  none when $\theta \neq 0$
  \\ 
  \hline
  \rule{0em}{1.5em} \makecell[r]{Additional \\ Low-Energy Solutions} & at $\vp\,\delta x=(\pm\pi,\pm\pi,\pm\pi)$ & \makecell{at $\vp = \pm(q,q,q)$ \\ where $\tan(q\delta x/2) = (\cos(\theta)+\sin(\theta))^{-1}$}
  \\
  \hline
   Range of Eigenvalues & $\abs{E\delta t} \leq \pi$ & $\abs{E\delta t} \leq 3(\pi - 2\theta) + mc^2\delta t$
   \\
   \hline
\end{tabular}
    \caption{Comparison between the conventional and the new Dirac walks in 3+1-D. The new walk is rather a family of walks parametrised by $\theta \in (-\pi/2,\pi/2)$. When $\theta=0$, the new walk reduces to the conventional walk. 
    Introducing the parameter $\theta$ allows us to get rid of any doublers or pseudo-doublers.}
    \label{tbl:3d}
\end{table*}

\subsection{3+1-D}

Again, we start with 

\begin{align}
    K_j =  \gamma^j_{+} S_j + \gamma^j_{0}+ \gamma^j_{-} S_j^\dagger
    \tag{\ref{eq:KjAbstract}}
\end{align}

Earlier, we had pick $\gamma^j_{0}=0$ to get the conventional Dirac walk in 3+1-D. However, now, we wish there to be some non-zero probability, in general, for the particle to remain where it is. Thus, we shall pick $\{\gamma^j_\mu\}$ more generally.

To this goal, similar to the 1-D case, let $A_{j},B_{j}$ be some subspaces of $\Hc_{\text{S}}$. Also, let $\Pi^{j}_{a}$ and $\Pi^{j}_{b}$ be projector operators into subspaces $A_{j}$ and $B_{j}$ respectively. We can also define the complimentary projector operators, $\Pi^{j}_{\bar{a}},\Pi^{j}_{\bar{b}}$, similar to the 1+1-D case. Then, we pick
\begin{align}
    \gamma^{j}_{+} &= \Pi^{j}_{a}\Pi^{j}_{b}\,,
    \nonumber\\
    \gamma^{j}_{-} &= \Pi^{j}_{\bar{a}}\Pi^{j}_{\bar{b}}\,,\,\text{and}
    \nonumber\\
    \gamma^{j}_{0} &= 1 - \gamma^{j}_{+} - \gamma^{j}_{-} = (\Pi^{j}_{a}\Pi^{j}_{\bar{b}}+\Pi^{j}_{\bar{a}}\Pi^{j}_{{b}}).
\end{align}
It can be verified that $K_j$ is indeed unitary for all $j$.

Now, let us consider the Weyl walks: $K^{+} = K_zK_yK_x$ and ${K}^{-}=K_z^\dagger K_y^\dagger K_x^\dagger$. Also, in the momentum space, ${K}^{+}(\vec{p}\,)=K^{-}(-\vec{p}\,)$. In the continuum limit, computed in a similar manner as before, we have
\be
    H^{\pm}_{\text{eff}} = \pm\sum_j (\gamma^j_{+}-\gamma^j_{-}) = \pm\sum_j (\Pi^{j}_{a}\Pi^{j}_{b} -\Pi^{j}_{\bar{a}}\Pi^{j}_{\bar{b}} )P_j v 
\ee
where, again, $v = \delta x/ \delta t$.

Let us now specify $A_j$ and $B_j$. Again, we introduce the parameter $-\pi/2< \theta <\pi/2$. Then, let us define the following operators.
\begin{align}
     \sigma_\theta^x &= \cos(\theta)\sigma_x - \sin(\theta)\sigma_z 
     \nonumber\\
     \sigma_\theta^y &= \cos(\theta)\sigma_y - \sin(\theta)\sigma_x
     \nonumber\\
     \sigma_\theta^z &= \cos(\theta)\sigma_z - \sin(\theta)\sigma_y 
     \label{eq:abchoice3d}
\end{align}

Then, we consider the eigenstates of these operators, $\sigma_\theta^j\ket{\uparrow^j_\theta}=\ket{\uparrow^j_\theta}$. Choosing  $\Pi^j_a = \ketbra{\uparrow^j_\theta}$ and $\Pi^j_b = \ketbra{\uparrow^j_{-\theta}}$, we get

 \begin{figure}
        \centering
		\includegraphics[width=0.9\linewidth]{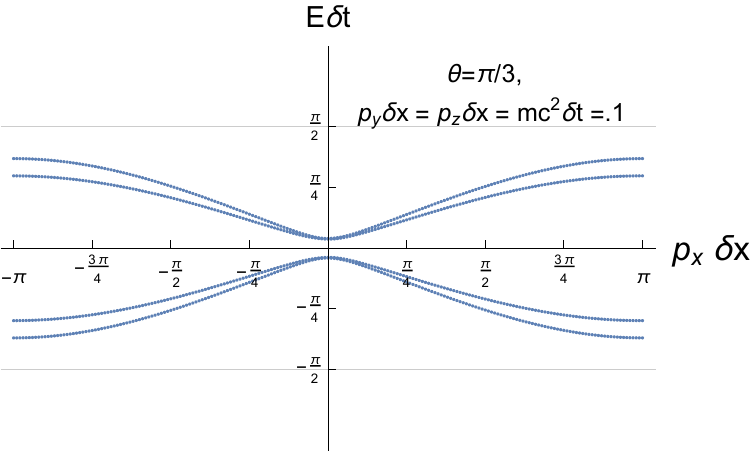}
        \caption{Plot of the dispersion relation for the parametrised 3+1-D Dirac quantum walk with $\theta=\pi/3$. We can see that the energy is never greater than $\pi/(2\,\delta t)$.}
        \label{fig:3DGraphTheta}
\end{figure} 

\be
    H^{\pm}_{\text{eff}} = \pm\vec{\sigma}\cdot\vec{P}v\cos(\theta)
\ee
which gives us the Weyl Hamiltonian if we take $c = v\cos(\theta)$. 

Now, consider a particle with momentum near $\vec{q} = (q,q,q)^{\bf{T}}$ where\footnote{
    As $q \delta x \in (-\pi,\pi)$, Eq.~\eqref{eq:samepoint} gives a unique $q$ for each $\theta$.
}
\be
    \tan(q \delta x/2) = \frac{1}{\cos(\theta)+\sin(\theta)}.
    \label{eq:samepoint}
\ee
We can show that for $\eta_j \delta x \ll 1$, 
\be
    K^{\pm}(\pm\vec{q}+\vec{\eta}\,) = I \mp i \,\delta x \cos(\theta) \vec{\eta}\cdot\vec{\sigma}^\prime + \Oc ((\eta_j \delta x)^2)
\ee
where $\{\sigma_j^\prime\}$ are an alternative representation of Pauli operators (Appendix \ref{app:fermiondoublingweyl}). Thus, we have fermion doubling at $\vp=\pm \vec{q}$ for the Weyl walks $K^\pm$ respectively (because the respective walkers behave like Weyl particles at these points). We have plotted the $p_x=p_y=p_z$ slice of the dispersion relation in Figure \ref{fig:samep} in which the doublers are noticeable.

Do we have any other doublers? For $\theta = 0$, we already listed the doublers exhaustively in Section \ref{sec:3dnormal}. Now, for $\theta \neq 0$, we expect there to be no other doublers apart from the one mentioned above. We performed a numerical search for a various values of $\theta$ to look for eigenvalues near zero, however we couldn't find any additional ones.

Now, again, we have Eqs.~\eqref{eq:mass3d}-\eqref{eq:3dwalkdefine} defining the Dirac walk as follows,
\be
    U = W  \begin{pmatrix}
        K_zK_yK_x & 0 \\
        0 & K_z^\dagger K_y^\dagger K_x^\dagger
    \end{pmatrix} = W \begin{pmatrix}
        K^{+} & 0 \\
        0 & K_{-}
    \end{pmatrix}.
    \tag{\ref{eq:3dwalkdefine}}
\ee

Then, for the Dirac walk, we have,
\begin{align}
 H_{\text{eff}} 
 &= 
 \begin{pmatrix}
     \sum_j(\Pi^{j}_{a}\Pi^{j}_{b} -\Pi^{j}_{\bar{a}}\Pi^{j}_{\bar{b}} )P_jv & \!\!\!mc^2 \\
    mc^2 & \!\!\!-\sum_j(\Pi^{j}_{a}\Pi^{j}_{b} -\Pi^{j}_{\bar{a}}\Pi^{j}_{\bar{b}} )P_jv
 \end{pmatrix}\!
\end{align}
in the continuum limit (when $p\delta x \ll 1$ and $mc^2\delta t \ll 1$). With the choice of $A^{j}$ and $B^{j}$, we get
 \begin{align}
     H_{\text{eff}} 
     &=
     \begin{pmatrix}
         \vec{\sigma}\cdot\vec{P}v\cos(\theta) & mc^2 \\
        mc^2 & -\vec{\sigma}\cdot\vec{P}v\cos(\theta)
     \end{pmatrix}
    \end{align}
in the continuum limit. Then, identifying $c = v \cos(\theta)$, we get the single particle 3+1-D Dirac Hamiltonian. 

We can again define a notion of energy for all momentum eigenstates (with any value of $mc^2 \delta t$). Note that because the internal state of the system is four-dimensional, for each $\vp\in\Theta$, we have four energy eigenvalues. We have plotted the dispersion relation for this walk for a particular value of (non-zero) $\theta$ in Figure \ref{fig:3DGraphTheta}.

We have already discussed fermion doubling for this walk when $\theta = 0$; for this choice of $\theta$, we get back the conventional 3+1-D Dirac walk. Let us discuss the case for non-trivial values of $\theta$. Because we have constructed the Dirac walk from the two Weyl walks, we expect to inherit any doublers. However,  for $\theta  \neq 0$ we do not have a good continuum limit at $\pm\vec{q}$. This is because $K^{\pm}$ have doublers at different points. These particles are still likely to cause issues in scattering models because two of the energy eigenvalues are zero. We do not expect introducing the mass term to produce more doublers. Thus, we expect there to be no doublers in this walk when $\theta \neq 0$.

Let us now make the case that we can always pick $\theta$ such that there are no pseudo-doublers. In Appendix \ref{app:choosetheta}, we show that
\be
    \abs{ E_{\vp}\,\delta t} \leq 3(\pi-2\theta)+ mc^2\delta t.
    \label{eq:choosethetamain}
\ee
Thus, for sufficiently small values of $mc^2\delta t$, we can always pick $\theta$ so that $\abs{ E_{\vp}\,\delta t}<\pi/2$. Because pseudo-doublers must have energy in a small ball around $\pm\pi/\delta t$, this ensures that we have no pseudo-doublers.

To conclude, for this choice of projectors, we have obtained a family of Dirac walks in 3+1-D parametrised by $\theta$. While, for $\theta  = 0$, we get back the conventional 3+1-D Dirac Walk, we can tune $\theta$ to remove all fermion doubling or pseudo-doubling. However, we do have extraneous low-energy solutions. Finally, for the 3+1-D case, we summarised all the differences between the family of walks and the conventional walk in TABLE \ref{tbl:3d}. 
 
\section{Discussion}

The QWs presented here can be second quantised to obtain their respective QCA versions (See Appendix \ref{app:QCAFree}). We can also add further interactions to such models. For instance, we provide an interacting model constructed from the family of quantum walks in Appendix \ref{app:QCAInt}. In particular, we introduce electric interactions in 1+1-D, thereby developing a QCA version of the Schwinger model by following Ref.~\cite{arrighi2020quantum}.

The notion of doublers and pseudo-doublers carries over to the respective QCA models. In particular, there are pseudo-doublers, particles that behave like low-momentum particles, but with energy near $\pm\pi/\delta t$ in the conventional Dirac QCA. We had pointed out in Ref. \cite{Gupta2025diracvacuumin} that these pseudo-doublers, for the conventional Dirac QCAs, are likely to make the vacuum state highly unstable if we were to introduce interactions.

For the Dirac field, we can define vacuum state by filling up all the negative energy states \cite{brun2025quantum}. All the filled negative energy states form what is called the Dirac sea. Unlike the continuum limit, where there is only one boundary between the Dirac sea and empty positive energy states, in the discrete theory, we have two boundaries between the Dirac sea and the positive energy states \textemdash $E=0$ and $E=\pm\pi/\delta t$. In the conventional Dirac QCA, if we were to introduce some interactions, we would expect high energy pair creation to take place near the $E=\pm\pi/\delta t$ boundary due to the presence of pseudo-doublers there. Moreover, particles with momentum close to $p_j=\pm \pi/(2\delta x)$ in 3+1-D, also have energy near both the boundaries, and thus may also contribute to the pair creation.

We can use the family of Dirac QCAs to solve this  problem. By choosing an appropriate value of $\theta$, we can make sure that there are no energy eigenvalues near $\pm\pi/\delta t$. Using the results about the eigenvalues of a product of unitaries discussed in Appendix \ref{app:product}, we show, in Appendix \ref{app:choosetheta}, that we can choose a $\theta$ such that $\abs{E_p\delta t} \leq \pi/2$ if given small enough values of mass. In Figure \ref{fig:1dplots} and Figure \ref{fig:3DGraphTheta}, one can see that the energy eigenvalues strictly lie between $-\pi/2\delta x$ and $\pi/2\delta x$. 

In our previous work, we had introduced a modified quantum walk that also resolved this issue. However, while this walk got rid of any pseudo-doublers and other particles near $E=\pm \pi/\delta t$, it had introduced more fermion doubling near $E=0$. On the contrary, the family of walks, that we have presented in this paper, are mostly free of both doublers and pseudo-doublers (for clever choices of $\theta \neq 0$).

Although, these walks are not free of all extraneous low-energy solutions. Considering the 3+1-D family of Dirac walks, for each value of $\theta$, note that we have have two extraneous low energy solutions ($\pm \vec{q}(\theta)$). Note that while we have constructed the 1+1-D walk in a very general manner, we have constructed the 3+1-D walk by simply combining the 1+1-D walks in each direction. We might be able to get rid of any extraneous low-energy solutions by considering a more general construction in higher dimensions. This could be an avenue to explore in a future work.

We wish to make a small note on the physical implementation of these walks. Suppose we have double the lattice sites and we always have an initial state on a superposition of odd sites. Consider the unitaries $V(\theta)=\ketbra{\uparrow_\theta}{\uparrow_z}+\ketbra{\downarrow_\theta}{\downarrow_z}$ and $T = \ketbra{\uparrow_z} S + \ketbra{\downarrow_z} S^\dagger$. Now, the following sequence of unitaries $V(\theta) T V(\theta)^\dagger V(-\theta) T V(-\theta)^\dagger$ implements our new walk with some additional lattice sites. Note  that the final state ( after a few applications of the above sequence of unitaries) will always be in a superposition of odd lattice sites. Ignoring the `temporary' even lattice sites for the initial and final states,  this decomposition consists of unitaries that have already been implemented in  existing experimental frameworks \cite{wang2013physical}, hence similar approaches could be used to implement our new walk. A similar decomposition can be achieved in higher dimensions.

Finally, we have based our argument on free theories. For future work, it would be interesting to explore scattering calculations in explicit interaction models \cite{destri1987light,eon2023relativistic}, including the one in Appendix \ref{app:QCAInt}, to have a closer look at the issues caused by the presence of these doublers and pseudo-doublers.

\section*{Acknowledgements}
We are thankful to Dogukan Bakircioglu and Pablo Arnault for an insightful discussion. 
\section*{DATA AVAILABILITY STATEMENT}
The code used to generate the plots in this paper and to analyse the dispersion relations is publicly accessible \cite{mydataset}.

\appendix

\section{Simulations}
\label{app:simulation}

\begin{figure*}
    \centering
    \includegraphics[width=0.8\linewidth]{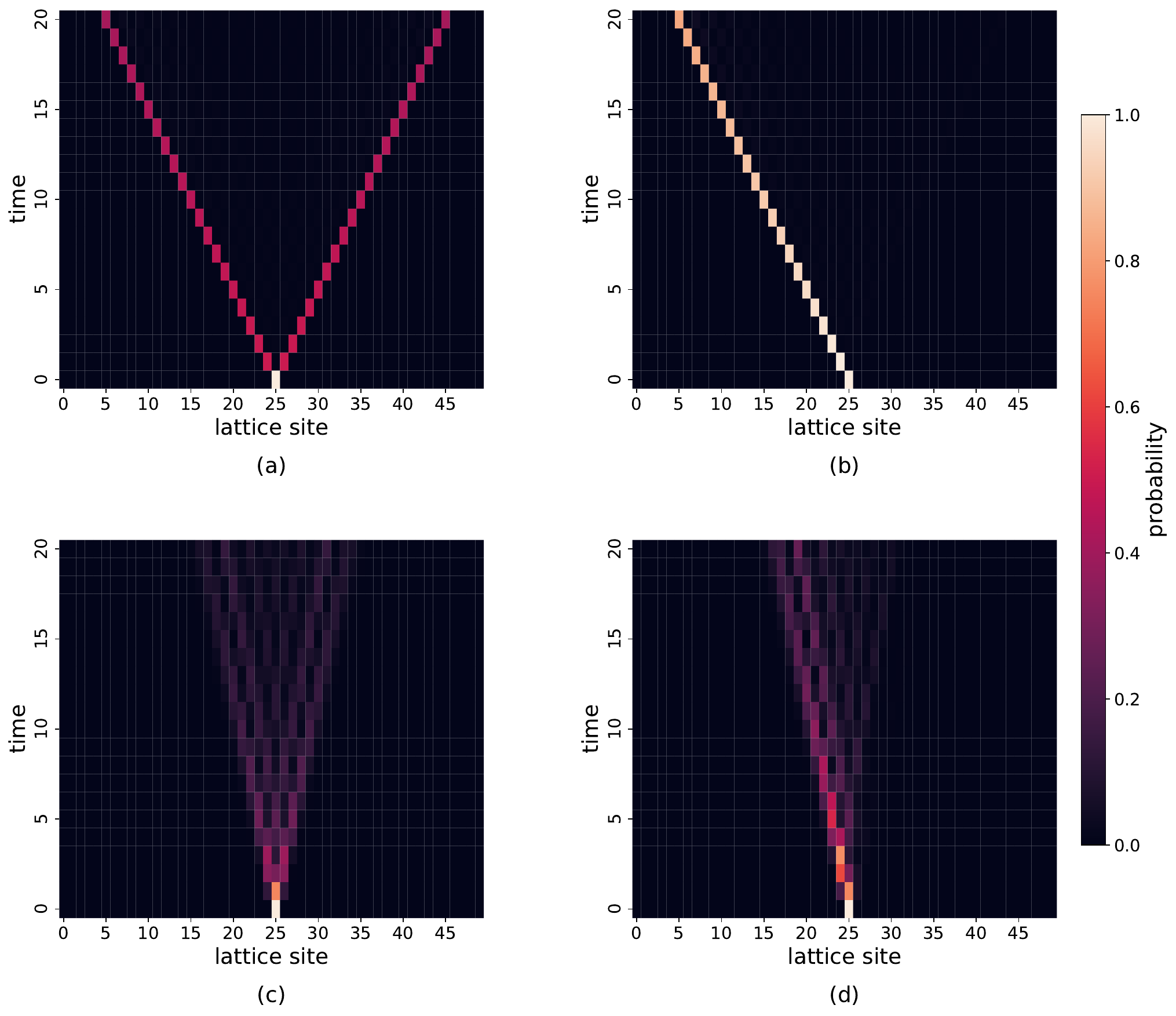}
    \caption{We have simulated the conventional and the new Dirac walks in 1+1-D and plotted  heatmaps of the probability of a walker being on a particular lattice site at some time. We considered a 1-D lattice consisting of 50 lattice sites and ran the simulation for 20 time steps. The initial state is taken to be a product state with the walker starting on the $\rm{25}^{\rm{th}}$ lattice site. For (a) and (b), we have considered the conventional walk ($\theta = 0$), and for (c) and (d), we have considered the new walk with $\theta=\pi/3$. The initial internal spin state is taken to be $(\ket{\uparrow_z}+\ket{\downarrow_z})/\sqrt{2}$ for (a) and (c), while it is taken to be $\ket{\downarrow_z}$ for (b) and (d). In all cases, we have taken $mc^2\delta t = 0.1$.}
    \label{fig:heatmap}
\end{figure*}

\begin{figure}[h!]
    \centering
    \includegraphics[width=0.95\linewidth]{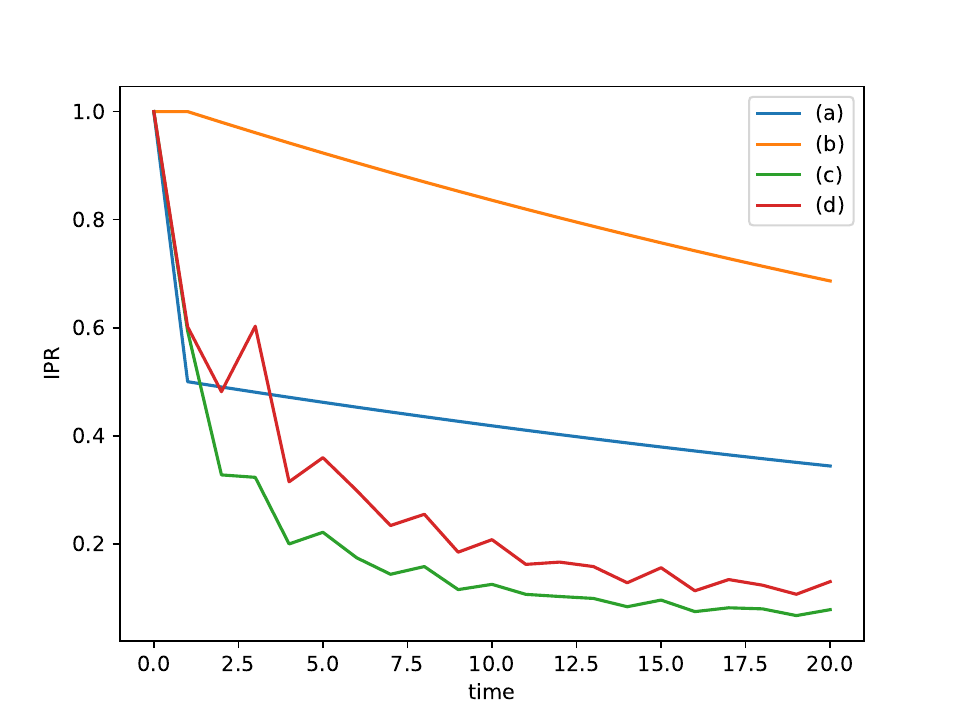}
    \caption{Plot of Inverse Participation Ratio (IPR) against time for the Dirac walks in 1+1-D. The IPR for a state $\ket{\psi(t)}$ is given by $\text{IPR}(t) = \sum_n \abs{\braket{\psi(t)}{n}\braket{n}{\psi(t)}}^2$. We consider an initial product state localised at a particular lattice site. For (a) and (b), we have considered the conventional walk ($\theta = 0$), and for (c) and (d), we have considered the new walk with $\theta=\pi/3$. The initial internal spin state is taken to be $(\ket{\uparrow_z}+\ket{\downarrow_z})/\sqrt{2}$ for (a) and (c), while it is taken to be $\ket{\downarrow_z}$ for (b) and (d). In all cases, we have taken $mc^2\delta t = 0.1$.}
    \label{fig:IRP}
\end{figure}

We ran simulations to compare the behaviour of the new and the conventional walks which are shown in Figures \ref{fig:heatmap} and \ref{fig:IRP}. There is a large distinction between the two walks as apparent from the Figures. However, we have simulated a state initially localised at a particular site which has high average energy, and thus is not in the regime where we would expect continuum behaviour.

Moreover, from the Figures, we can also see that we have sub-ballistic behaviour for the new walk. For very small mass ($mc^2 \delta t$), sub-ballistic behaviour is largely negligible in the conventional walk over the times considered. Furthermore, for the same underlying lattice and size of time step, the new walk is slower than the old walk. However, for the purpose of simulating field theory, to bring the new walk up to speed, we can rescale the underlying lattice accordingly as discussed earlier. In particular, for the old walk, we set $\delta x/\delta t = c$, and for the new walk we set $\delta x/\delta t = c/\cos(\theta)$ where $\delta x$ and $\delta t$ are the lattice spacing and size of time step respectively.

\section{Fermion Doubling in 3+1-D at \texorpdfstring{$p_j = \pm \pi/2\delta x$}{TEXT}}
\label{app:weirdpoint}
Consider the 3+1-D conventional Weyl walk from Section \ref{sec:3dnormal}. We have,
\be
    K_j( \pm \pi/2\delta x + \eta) = \mp i \sigma_j K_j (\eta)
    \label{eq:Kjweirdpointexpand}
\ee
with $K_j(p) = \Exp{-ip\sigma_j\delta x}$.
Power expanding for $\eta\delta x \ll 1$, we get
\begin{align}
    K_j(\pm \pi/ 2 \delta x +\eta) &= \mp i \sigma_j (I-i\eta\sigma_j\delta x)+\Oc(\delta x^2)
    \nonumber\\
    &= \mp (i \sigma_j + \eta\delta x)+\Oc(\delta x^2).
\end{align}

Then, for the Weyl walk, if $\vp=(\pi/2 \delta x,\pi/2 \delta x,\pi/2 \delta x)^\mathbf{T}$ or $\vp=(-\pi/2 \delta x,-\pi/2 \delta x,\pi/2 \delta x)^\mathbf{T}$ (including all the permutations), we have

\begin{align}
    K(\vp+\vec{\eta}\,)
    &= -(I + i \vec{\eta}\!\cdot\vec{\sigma}''\delta x)+\Oc(\delta x^2)
\end{align}
where $\vec{\sigma}\,''\ = (-\sigma_x,\sigma_y,-\sigma_z)^\mathbf{T}$ is an alternative representation of the Pauli matrices. Around this point, the particles behave like Weyl particles of the opposite chirality, and have energy near $\pm\pi/\delta t$. Thus, we could consider these points to be pseudo-doublers. 

We also have, for $\vp=(-\pi/2 \delta x,-\pi/2 \delta x,-\pi/2 \delta x)^\mathbf{T}$ or $\vp=(-\pi/2 \delta x,\pi/2 \delta x,\pi/2 \delta x)^\mathbf{T}$ (including all the permutations),
\begin{align}
    K(\vp+\vec{\eta}\,)
    &= (I + i \vec{\eta}\!\cdot\vec{\sigma}\,''\delta x)+\Oc(\delta x^2).
    \label{eq:weirdpseudodoubler}
\end{align}
Thus, these points behave like Weyl particles of opposite chirality, with  energy near $0$, and hence we could consider them doublers. 

Let us now look at the Dirac walk. With $\ket{\psi_p}$ as an energy eigenstate, for $\vp=(\pi/2 \delta x,\pi/2 \delta x,\pi/2 \delta x)^\mathbf{T}$, 

\begin{align}
    T\ket{\psi_{\vp+\vec{\eta}}}  &= \begin{pmatrix}
        K_zK_yK_x & 0 \\
        0 & K_z^\dagger K_y^\dagger K_x^\dagger
    \end{pmatrix}\!\ket{\psi_{\vp+\vec{\eta}}}
    \nonumber\\
    &\!\!\!\!\!\!\!\!\!\!\!\!\!\!\!\!\!\!= \left(\begin{pmatrix}\!
          -I+i\vec{\eta}\!\cdot\vec{\sigma}\,'\delta x  & 0 \\
        0 &  I+i\vec{\eta}\!\cdot\vec{\sigma}\,'\delta x
    \end{pmatrix}\!\!+\!\Oc(\delta x^2)\right)\!\ket{\psi_{\vp+\vec{\eta}}}.
    \nonumber\\
\end{align}
Then, with the mass term $W$, we have,
\begin{align}
    WT\ket{\psi_{\vp+\vec{\eta}}}  &= \Bigg[\begin{pmatrix}
          -I+i\vec{\eta}\!\cdot\vec{\sigma}\,'\delta x & -imc^2\delta t \\
        -imc^2\delta t &  I+ i\vec{\eta}\!\cdot\vec{\sigma}\,'\delta x
    \end{pmatrix}
    \nonumber\\
    &\quad\quad\quad+\Oc(\delta x^2,\delta x\delta t,\delta t^2)\Bigg]\!\ket{\psi_{\vp+\vec{\eta}}}.
\end{align}

We get a similar expression for the other points. As, we can't expand around identity around these points, i.e., $WT\ket{\psi_{\vp}}\not\approx e^{i\phi}I \ket{\psi_{\vp}}$, for any arbitrary phase $e^{i\phi}$, we do not have a good continuum limit around these point for the massive case.

\section{Unitarity of the family of Walks}
\label{app:unitarity}

From $    T = \gamma_{+} S + \gamma_{0} + \gamma_{-} S^\dagger$, we get
\begin{align}
    T^\dagger T &= (\gamma^\dagger_{+}S^\dagger+\gamma_0^\dagger + \gamma_{-}^\dagger S)(\gamma_{+}S+\gamma_0 + \gamma_{-} S^\dagger)
    \nonumber\\
    &= (\gamma^\dagger_{+}\gamma_{-})S^{\dagger2}+(\gamma_{+}^\dagger\gamma_0+\gamma_0^\dagger\gamma_{-})S^\dagger
    \nonumber\\
    &+ (\gamma_{+}^\dagger\gamma_{+}+\gamma_{0}^\dagger\gamma_{0}+\gamma_{-}^\dagger\gamma_{-})
    \nonumber\\
    &+ (\gamma_{0}^\dagger\gamma_{+}+\gamma_{-}^\dagger\gamma_{0})S +(\gamma^\dagger_{-}\gamma_{+})S^{2}.
    \label{eq:app:expand}
\end{align}

Recalling that 
\begin{align}
    \gamma_{+} &= \Pi_{a}\Pi_{b}
    \nonumber\\
    \gamma_{-} &= \Pi_{\bar{a}}\Pi_{\bar{b}}
    \nonumber\\
    \gamma_{0} &  = (\Pi_{a}\Pi_{\bar{b}}+\Pi_{\bar{a}}\Pi_{{b}})
    \tag{\ref{eq:gammaValue}},
\end{align}
we will now prove the unitarity of $T$. We will use the following identities in this section: $\Pi_a+\Pi_{\bar{a}}=I$ and $\Pi_{\bar{a}}\Pi_a=\Pi_a\Pi_{\bar{a}}=0$.

Consider
\begin{subequations}
    \begin{align}
    \gamma_{+}^\dagger\gamma_{0} &= \Pi_b\Pi_a (\Pi_{\bar{a}}\Pi_b+\Pi_a\Pi_{\bar{b}}) = \Pi_b\Pi_a\Pi_{\bar{b}}
    \label{eq:app:term2}
    \\
    \gamma_{0}^\dagger\gamma_{-} &=  (\Pi_b\Pi_{\bar{a}}+\Pi_{\bar{b}}\Pi_a)\Pi_{\bar{a}}\Pi_{\bar{b}} = \Pi_b\Pi_{\bar{a}}\Pi_{\bar{b}}.
    \label{eq:app:term3}
\end{align}\label{eq:app:s2}
\end{subequations}
Then, adding Eqs.~\eqref{eq:app:term2} and \eqref{eq:app:term3}, we get
\be
     \gamma_{+}^\dagger\gamma_{0}+\gamma_{0}^\dagger\gamma_{-}=
        \Pi_b\underbrace{(\Pi_a+\Pi_{\bar{a}})}_{=I}\Pi_{\bar{b}} = \Pi_b\Pi_{\bar{b}} = 0 
        \label{eq:app:term1}.
\ee

Also, consider,
\be
    \gamma_{+}^\dagger\gamma_{-} = (\Pi_{a}\Pi_{b})^\dagger (\Pi_{\bar{a}}\Pi_{\bar{b}})
    = \Pi_{b}\underbrace{\Pi_{a}\Pi_{\bar{a}}}_{=0}\Pi_{\bar{b}}=0.
    \label{eq:app:term4}
\ee

Because of the symmetry between the barred and unbarred projection operators, from Eqs.~\eqref{eq:app:term1} and \eqref{eq:app:term4}, we also have,
\begin{align}
    \gamma_{-}^\dagger\gamma_{+} = 0\,\text{, and}\label{eq:app:term5}\\
    \gamma_{0}^\dagger\gamma_{+}+\gamma_{-}^\dagger\gamma_{0}=0.
    \label{eq:app:term6}
\end{align}

Moreover, we get
\begin{align}
    \gamma_{+}^\dagger\gamma_{+}+\gamma_{-}^\dagger\gamma_{-}&=\Pi_b \Pi_a \Pi_b +\Pi_{\bar{b}} \Pi_{\bar{a}} \Pi_{\bar{b}} \,
    \label{eq:app:one}
    \text{, and} \\
    \gamma_{0}^\dagger\gamma_{0} &= \Pi_{\bar{b}} \Pi_a \Pi_{\bar{b}} +\Pi_b \Pi_{\bar{a}} \Pi_b 
    \label{eq:app:two}
\end{align}
Thus, after adding Eqs.~\eqref{eq:app:one} and \eqref{eq:app:two}, we get
\begin{align}
    \gamma_{+}^\dagger\gamma_{+}+\gamma_{-}^\dagger\gamma_{-}+\gamma_{0}^\dagger\gamma_{0}&=
    \Pi_{\bar{b}}(\underbrace{\Pi_a+\Pi_{\bar{a}}}_{I})\Pi_{\bar{b}}
    \nonumber\\
    &\quad+\Pi_b(\underbrace{\Pi_a+\Pi_{\bar{a}}}_{I})\Pi_b
    \nonumber\\
    &=\Pi_{\bar{b}}+\Pi_{b}=I.
    \label{eq:app:some}
\end{align}

Finally, using Eqs.~\eqref{eq:app:term1}-\eqref{eq:app:term6} and Eq.~\eqref{eq:app:some} in Eq.~\eqref{eq:app:expand}, we get $T^\dagger T = I$. In a similar way, we can also show that $ TT^\dagger = I$.

\section{Behaviour of the 1+1-D walker in a small ball around any momentum}
\label{app:continuum}
Consider the family of quantum walks from \ref{sec:slowwalk1d}. We shall look at the case when $m=0$. We can rewrite $T$ from Eq.~\eqref{eq:unitaryabstract2} as
\begin{align}
    T &= \Exp{-iP(\Pi_{a}-\Pi_{\bar{a}})\delta 
    x/2}
    \nonumber \\ &\quad \times
    \Exp{-iP(\Pi_{b}-\Pi_{\bar{b}})\delta x/2}.
\end{align}

With the choice of $\Pi_a = \ketbra{\uparrow_\theta}{\uparrow_\theta}$ and $\Pi_b = \ketbra{\uparrow_{-\theta}}{\uparrow_{-\theta}}$ as before, we can write $T$ as
\begin{align}
    T = \Exp{-iP\sigma_\theta\delta 
    x/2}
    \Exp{-iP\sigma_{-\theta}\delta x/2}.
\end{align}.

Acting $T$ on a momentum eigenstate with momentum $p\,\delta x\in(-\pi,\pi]$, we can define $T(p)$ as
\be
      T(p) = \Exp{-ip\sigma_\theta\delta 
    x/2}
    \Exp{-ip\sigma_{-\theta}\delta x/2}.
    \label{eq:exactTp}
\ee

Further simplifying, we have
\begin{align}
    T(p) &= 
    (\cos(p\delta x/2)-i\sigma_\theta\sin(p\delta x/2))
    \nonumber\\ &\quad\times(\cos(p\delta x/2)-i\sigma_{-\theta}\sin(p\delta x/2)) 
    \nonumber\\
    &= \cos^2(p\delta x/2)- i\cos(p\delta x/2)\sin(p\delta x/2)(\sigma_\theta+\sigma_{-\theta})\nonumber\\
    &\quad-\sigma_\theta\sigma_{-\theta}\sin^2(p\delta x/2)
    \nonumber\\
    &= \cos^2(p\delta x/2)- i\sin(p\delta x)\cos(\theta)\sigma_z \nonumber\\ \quad &\quad\quad- e^{-i2\theta\sigma_x } \sin^2(p\delta x/2).
    \label{eq:Tpfinal}
\end{align}

Now, to have a good continuum limit near $p$, we require that for $\eta \delta x \ll 1$,
\be
    T(p+\eta)=e^{i\phi}(I - iH\eta + \Oc(\eta^2))
\ee
for some $\phi$ independent of $\eta$ and $H$ Hermitian.

Thus, we want $T(p)=e^{i\phi}I$ for some $\phi$. Now, as we are dealing with unitaries that are in $SU(2)$, the only possibilities are $e^{i\phi} = \pm 1$. From Eq.~\eqref{eq:Tpfinal}, and recalling that $\pi/2 < \theta < - \pi/2$, we can see that this is only possible when either (a) $p= 0$ with any $\theta$ or (b) $p=\pi/ \delta x$ with $\theta = 0$. We already know that these points indeed give a good continuum limit. The same line of reasoning holds if we include the mass term.

\section{Eigenvalues of Product of Unitaries}
\label{app:product}
Suppose we have finite-dimensional unitary matrices $U$ and $V$. Then we can always write $U={e^{iH}}$ and $V=e^{iG}$ where $H$ and $G$ are Hermitian. We can always pick $H$ and $G$ such that their eigenvalues lie in $(-\pi,\pi]$. Let the eigenvalues of $H$ and $G$ be $\{\lambda_{k}\}_{k}$ and $\{\eta_{k}\}_{k}$. Let $\lambda$ and $\eta$ be non-negative real numbers such that $\abs{\lambda_i}\leq \lambda \leq \pi$ and $\abs{\eta_i}\leq \eta \leq \pi$ for all $i$.

By Thompson's theorem \cite{thompson1974eigenvalues}, we have
\be
    UV=e^{i(H+W G W^\dagger)}
\ee
where $W$ is some unitary matrix. Let the eigenvalues of $H+W G W^\dagger$ be $\{\gamma^\prime_k\}_{k=1}^{d}$. 

Then applying Weyl's inequality \cite{horn1962eigenvalues}, we get,
\begin{subequations}
    \begin{align}
        \max_k \gamma^\prime_{k} \leq \max_k \lambda_k + \max_k \eta_k \leq \lambda + \eta, \\
        \min_k \gamma^\prime_{k} \geq - \min_k \lambda_k - \min_k \eta_k \geq -\lambda - \eta.
    \label{eq:linearlagebrainterim}
    \end{align}
\end{subequations}

In other words, for all $k$,
\be
    \abs{\gamma^\prime_k} \leq \abs{\eta}+ \abs{\lambda}.
    \label{eq:linearalgebramain1}
\ee
    
Now, let the eigenvalues of $UV$ be $\{e^{-i\gamma_k}\}_k$ such that $\gamma_k\in(-\pi,\pi]$. We can obtain $\{\gamma_k\}_k$ from $\{\gamma^\prime_k\}_k$ by shifting each $\gamma^\prime_k$ individually by an integral number of $2\pi$ till it is within $(-\pi,\pi]$. Thus, from Eq.~\eqref{eq:linearalgebramain1}, it follows that
\be
    \abs{\gamma_k} \leq \abs{\gamma^\prime_k} \leq \abs{\eta}+ \abs{\lambda}.
    \label{eq:linearalgebramain}
\ee

\section{Restricting Energy Eigenvalues}
\label{app:choosetheta}

Consider $K_j$ used to define the family of walks from section \ref{sec:slowwalk}. As $K_j$ is translationally invariant, in momentum space, we write it as a unitary operator $K_j(p)$ acting on $\Hc_{\text{S}}$. Now, we wish to write an explicit expression for the real part of the eigenvalue of $K_j(p)$. 

From Eq.~\eqref{eq:exactTp}, we know that we can write $K_j(p) = rI + i A$ where A is a Hermitian matrix with
\be
    r =  \cos^2(p\delta x/2)-\sin^2(p\delta x/2)\cos(2\theta).
\ee
Then, $r$ is indeed the real part of the eigenvalues of $K_j(p)$.  We have the same expression for all $j$'s because of the symmetry between different $K_j$. 

 Now, because $K_j(p)$ is unitarily equivalent to its Hermitian conjugate, we can let its eigenvalues to be $e^{\pm i\lambda_j}$ with $0 \leq \lambda_j \leq \pi$. Then, we can write
\be
    \cos(\lambda_j) = \cos^2(p\delta x/2)-\sin^2(p\delta x/2)\cos(2\theta).
\ee

Now, we wish to find the minimum of this expression for $p\,\delta x \in (-\pi,\pi]$. Recall that $-\pi/2<\theta<\pi/2$. We shall further impose that $0\leq\theta<\pi/2$. We have,
\be
    \min_{-\pi<p\leq\pi}\cos(\lambda_j) = -\cos(2\theta)=\cos(\pi-2\theta).
\ee
Because $0< \pi-2\theta \leq \pi $ and $0 \leq \lambda_j \leq \pi$, we get, for all $j$,
\be
    \lambda_j \leq \pi-2\theta.
\ee

Now, consider $T_j$. Acting them on momentum eigenstates, we can write them as $T_j(p)$. The eigenvalues of $K_j$ and $T_j$ are the same by construction. Also note that the eigenvalues of $W$ are $\pm mc^2 \delta t$ with $0\leq mc^2 \delta t\leq\pi$. Then, using the result of the previous section (Eq.~\eqref{eq:linearalgebramain}), we have
\be
    \abs{ E_{\vp}\,\delta t} \leq 3(\pi-2\theta)+ mc^2\delta t.
    \label{eq:choosetheta}
\ee

Finally, we can see from Eq.~\eqref{eq:choosetheta}, for all $0\leq mc^2\delta t < \pi/2$, we can always pick a $\theta$ with $0 \leq \theta < \pi/2$ such that $\abs{E_{\vp}\, \delta t}<\pi/2$.

In a similar manner, we can show that the above also holds true for the 1+1-D family of walks\footnote{For 1+1-D family of walks, we have 
    \be
    \abs{ E_p\delta t} \leq (\pi-2\theta)+ mc^2\delta t.
    \ee
Again, the same conclusion follows.
}.

\section{Fermion Doubling in the Weyl Walks}
\label{app:fermiondoublingweyl}

We shall consider the family of Weyl walks, $K^{+}= K_z K_y K_x$ and $K^{-}= K_z^\dagger K_y^\dagger K_x^\dagger$ from section \ref{sec:slowwalk} at the point $\vec{q}=q(1,1,1)^\mathbf{T}$ where $q$ is given implicitly by 
\be
    \tan(q(\theta) \delta x/2) = \frac{1}{\cos(\theta)+\sin(\theta)}.
    \tag{\ref{eq:samepoint}}
\ee

First, note that from Eq.~\eqref{eq:Tpfinal}, we can write $K_z(p)$ as
\begin{align}
    K_z(p) &= \cos^2(p\delta x/2)- i\sin(p\delta x)\cos(\theta)\sigma_z \nonumber\\ \quad &\quad\quad- e^{-i2\theta\sigma_x } \sin^2(p\delta x/2).
\end{align}

Then, after some algebra, we can write $K_z(q(\theta))$ as
\begin{align}
    K_z(q) &= \frac{1}{1+cs}\left( s(s+c) - i c (s+c) \sigma_z + i sc\sigma_x \right).
    \label{eq:samepKexplicit1}
\end{align}
where $s=\sin(\theta)$ and $c=\cos(\theta)$.

We can similarly write $K_y(q(\theta))$ and $K_x(q(\theta))$.
\begin{subequations}
    \begin{align}
    K_y(q) &= \frac{1}{1+cs}\left( s(s+c) - i c (s+c) \sigma_y + i sc\sigma_z \right).
    \\
    K_x(q) &= \frac{1}{1+cs}\left( s(s+c) - i c (s+c) \sigma_x + i sc\sigma_y \right).
\end{align}
    \label{eq:samepKexplicit2}
\end{subequations}

Then, we can verify that for any $\theta$,

\be
K^{+}(\vec{q}(\theta))=K_z(q) K_y(q) K_x(q)=I.
\ee

Thus, we do have a good continuum limit at this point. 

Let us now find the effective Hamiltonian. 
With
\be
      K_j = \Exp{-iP_j\sigma^j_\theta\delta 
    x/2}
    \Exp{-iP_j\sigma^j_{-\theta}\delta x/2},
\ee
we get, for $\eta_j \delta x \ll 1$,
\be
    K_j(p_j+\eta_j) \approx K_j(p_j) - \frac{i}{2}\eta_j\delta x(\sigma_{\theta}^j T_j(p_j) + T_j(p_j) \sigma_{-\theta}^j)
\ee
upto first order.

Then, for $K(\vec{p}+\vec{\eta}\,)$, we get
\begin{align}
    K^+(\vec{p}+\vec{\eta}) &\approx K_z(p_x)K_y(p_y)K_x(p_z)
    \nonumber
    \\
    &- \frac{i}{2} \eta_x \delta x K_z(p_z) K_y(p_y) (\sigma_{\theta}^x K_x(p_x) + K_x(p_x) \sigma_{-\theta}^x)
    \nonumber
    \\
    &- \frac{i}{2} \eta_y \delta x K_z(p_z)(\sigma_{\theta}^y K_y(p_y) + K_y(p_y) \sigma_{-\theta}^y) K_x(p_x)
    \nonumber
    \\
    &- \frac{i}{2} \eta_z \delta x (\sigma_{\theta}^z K_z(p_z) + K_z(p_z) \sigma_{-\theta}^z) K_y(p_y) K_x(p_x).
\end{align}

Now, consider $\vec{p} = \vec{q}(\theta)$. Using $K_z (q) K_y (q) K_x (q) = I$, we get

\begin{align}
    K^+(\vec{q}+\vec{\eta}) \approx I &- \frac{i}{2} \eta_x \delta x  (K_x^\dagger(q) \sigma_{\theta}^x K_x(q) +  \sigma_{-\theta}^x)
    \nonumber
    \\
    &- \frac{i}{2} \eta_y \delta x (K_z(q)\sigma_{\theta}^y K_z^\dagger(q) 
    \nonumber \\
    &\quad\quad\quad\quad+ K^\dagger_x(q) \sigma_{-\theta}^y K_x(q)) 
    \nonumber
    \\
    &- \frac{i}{2} \eta_z \delta x (\sigma_{\theta}^z + K_z(q) \sigma_{-\theta}^z K_z^\dagger(q)).
\end{align}

We can rewrite the above equation, compactly, as 
\be
    K^+(\vec{q}+\vec{\eta}) \approx I
    - i \cos(\theta) \vec{\eta}\cdot\vec{\sigma}^\prime \delta x
    \label{eq:finalsamep}
\ee
where 
\begin{align}
    \sigma^\prime_x &= (K_x^\dagger(q) \sigma_{\theta}^x K_x(q) +  \sigma_{-\theta}^x)/\cos(\theta),
    \nonumber \\
    \sigma^\prime_y &= (K_z(q)\sigma_{\theta}^y K_z^\dagger(q) + K^\dagger_x(q) \sigma_{-\theta}^y K_x(q))/\cos(\theta),
    \nonumber\\
    \sigma^\prime_z &= (\sigma_{\theta}^z + K_z(q) \sigma_{-\theta}^z K_z^\dagger(q))/\cos(\theta).
    \label{eq:rotatedPauli}
\end{align}

Using a computer algebra system, we verified that $[\sigma^\prime_i,\sigma^\prime_j]=i\epsilon_{ijk}\sigma^\prime_k$ and $\sigma_j^2 = I$. Thus, $\{\sigma^\prime_j\}$ are an alternative representation of Pauli operators.

Now, consider $K^{-}$. Because, $K^{-}(p)=K^{+}(-p)$, we have,
\be
    K^{-}(-\vec{q}+\vec{\eta}) \approx I
    + i \cos(\theta) \vec{\eta}\cdot\vec{\sigma}^\prime \delta x.
    \label{eq:finalsamep2}
\ee

Thus, we can see from Eqs.~\eqref{eq:finalsamep} and \eqref{eq:finalsamep2}, that the walkers given by $K^{\pm}$ behave like a Weyl particle at $\pm q$ respectively. Thus, we have fermion doubling at this point.

Are there any other doublers for the Weyl walks $K^{\pm}$? We attempted a numerical search for zero energy eigenvalues for different values of $\theta \neq 0$, and couldn't find any other low-energy solutions. Moreover, note that, from the Eq.~\eqref{eq:choosetheta}, it is evident that, we can choose $\theta$ such that $E_{\vec{p}}\, \delta t < \pi/2$. Thus, there are no pseudo-doublers in these walks (as the energy of a pseudo-doubler has to be $\pi/\delta t$).

\section{Free Dirac Quantum Cellular Automata in 1+1-D}
\label{app:QCAFree}
In this section, we shall construct a QCA which yields the family of Dirac walks in its single particle limit. Again, we suppose a 1-D lattice with spacing $\delta x$. Suppose we have a two component field on the lattice;
\be
    \PsiB{n} = 
    \begin{pmatrix}
        \psiF{n}{r}\\[0.1cm]
        \psiF{n}{l}
    \end{pmatrix}
\ee
such that $\{\psiDag{n}{a},\psiF{m}{b}\}=\delta_{nm}^{ab}$ where $n,m\in\Zb$ and $a,b\in\{r,l\}$.

Suppose $A,B$ are vector subspaces of $\mathbb{C}^2$ . Let $\Pi_a$ and $\Pi_b$ be projection matrices into $A$ and $B$ respectively.

We can implicitly define the unitary $T$ such that
\begin{align}
    T \PsiB{n} T^\dagger &= \Pi_a\Pi_b\PsiB{n+1}+(\Pi_{a}\Pi_{\bar{b}}
    \nonumber\\
    &+\Pi_{\bar{a}}\Pi_{{b}})\PsiB{n}+\Pi_{\bar{a}}\Pi_{\bar{b}}\PsiB{n-1}.
    \label{eq:TQCA}
\end{align}
More explicitly, we can write the unitary as
\begin{align}
    T &= \Exp{-i\,\frac{\delta x}{2}\!\int_{-\pi/\delta x}^{\pi/\delta x}dp \,p\, \PsiB{p}^\dagger\, (\Pi_a-\Pi_{\bar{a}})\PsiB{p}}
    \nonumber\\
    &\quad\times\Exp{-i\,\frac{\delta x}{2}\!\int_{-\pi/\delta x}^{\pi/\delta x}dp \,p\, \PsiB{p}^\dagger\, (\Pi_b-\Pi_{\bar{b}})\PsiB{p}}.
\end{align}
We can also define an additional unitary $W$ such that
\be
     W\begin{pmatrix}
       \psiF{n}{r}\\[0.1cm]
        \psiF{n}{l}
    \end{pmatrix}W^\dagger
    =
    \begin{pmatrix}
        W\psiF{n}{r}W^\dagger\\[0.1cm]
        W\psiF{n}{l}W^\dagger
    \end{pmatrix}
    =
   e^{-i M \delta t} 
   \begin{pmatrix}
        \psiF{n}{r}\\[0.1cm]
        \psiF{n}{l}
    \end{pmatrix}
\ee

where $M$ is a complex 2 by 2 Hermitian matrix.

Then, we can define a QCA via the unitary
\be
    U = WT.
\ee

Now, we shall identify $\ket{\uparrow_z}$ with the vector $(1,0)^\mathbf{T}$, and identify $\ket{\downarrow_z}$ with the vector $(0,1)^\mathbf{T}$.

Then, with
\be
    \Pi_{a} = \Pi_{b} = 
    \begin{pmatrix}
        1 & 0 \\
        0 & 0 
    \end{pmatrix},
\ee
we get,
    \be
     T\begin{pmatrix}
       \psiF{n}{r}\\[0.1cm]
        \psiF{n}{l}
    \end{pmatrix}T^\dagger
    =
    \begin{pmatrix}
        T\psiF{n}{r}T^\dagger\\[0.1cm]
        T\psiF{n}{l}T^\dagger
    \end{pmatrix}
    = 
   \begin{pmatrix}
        \psiF{n+1}{r}\\[0.1cm]
        \psiF{n-1}{l}
    \end{pmatrix}.
\ee

Alternatively, we can write
\begin{align}
    WT &= \Exp{-i\delta t \,\! \sum_{n\in\Zb}\PsiB{n}^\dagger\,M \PsiB{n}}
   \nonumber\\
    &\times
    \Exp{-i\,\delta x\!\int_{-\pi/\delta x}^{\pi/\delta x}dp \,p\, \PsiB{p}^\dagger\, Z\PsiB{p}}
\end{align}
where $Z$ is the Pauli-Z matrix. Now, we if see $M= mc^2X$ with $X$ being the Pauli-X matrix, we get the original Dirac QCA from Ref.~\cite{farrelly2014causal}.

Now, with the same identification as before, we identify the vectors
\be
\begin{pmatrix}
    \cos(\theta/2) \\
    -i\sin(\theta/2)
\end{pmatrix}
\text{  and   }
\begin{pmatrix}
    -i \sin(\theta/2) \\
    \cos(\theta/2)
\end{pmatrix}
\ee
with $\ket{\uparrow_\theta}$ and $\ket{\downarrow_\theta}$.
Let
\be
    \Pi_{\theta} = \frac{1}{2}
    \begin{pmatrix}
        1+\cos(\theta) & i\sin(\theta) \\
        -i\sin(\theta) & 1-\cos(\theta)
    \end{pmatrix}.
\ee
Then, with $\Pi_a = \Pi_\theta$ and $\Pi_b = \Pi_{-\theta}$, we get a family of Dirac QCAs in 1+1-D as follows,
\begin{align}
    WT &= \Exp{-i\delta t \,\! \sum_{n\in\Zb}\PsiB{n}^\dagger\,M \PsiB{n}}
   \nonumber\\
    &\times
    \Exp{-i\,\frac{\delta x}{2}\!\int_{-\pi/\delta x}^{\pi/\delta x}dp \,\PsiB{p}^\dagger\,p\,(\cos(\theta)Z-\sin(\theta)Y) \PsiB{p}}
    \nonumber\\
    &\times
    \Exp{ -i\,\frac{\delta x}{2} \!\int_{-\pi/\delta x}^{\pi/\delta x}\!\!\!\!dp\,\PsiB{p}^{\dagger}\,p\,(\cos(\theta)Z+\sin(\theta)Y)\PsiB{p} }.
\end{align}

We also set $M = mc^2X$ with $c= \cos(\theta)\tfrac{\delta x}{\delta t}$, to get the correct continuum limit.

Finally, note that the explicit construction of the 3+1-D free QCAs, obtained by `second-quantising' the 3+1-D Dirac walks, can be done in a similar manner to the 1+1-D case.

\section{Interacting Quantum Cellular Automata: Schwinger Model}
\label{app:QCAInt}

We shall now the use the framework of Ref.~\cite{arrighi2020quantum} to introduce electromagnetic interactions to the Dirac QCAs in 1+1-D. We expect the model to describe quantum electrodynamics (QED) in 1+1-D, also known as the Schwinger model.

Let us start with the model in the previous section. Let us define the unitary $G_{\alpha}$ where $\alpha: \Zb\to [0,2\pi)$ to implement gauge transformations as follows,
\be
    G_\alpha^\dagger \psi_n^{a}G_\alpha = \psi_n^{a} e^{i\alpha_n}.
    \label{eq:gauge1}
\ee
Clearly, $T$ as defined in Eq.~\eqref{eq:TQCA} is not gauge invariant, i.e., $[G_\alpha,T]\neq 0$ in general. 

We wish to construct a unitary that is gauge invariant. To this goal, we introduce a link between each consecutive lattice sites labelled by $n\in\Zb+1/2$. At each of these links, we introduce a Hilbert space spanned by orthonormal basis $\{\ket{l}_n, l \in \Zb $\}. We define the operators $E_n$ and $V_n$ implicitly via $E_n\ket{l}_n = l\ket{l}_n$ and $[V_n,E_n]=V_n$. Note that the operators on different lattice sites commute with each other. 

We define the gauge transformation on $\ket{l}_n$ as
\be
    G_\alpha\ket{l}_n = \Exp{il(\alpha_{n+1/2}-\alpha_{n-1/2})}\ket{l}_n.
    \label{eq:gauge2}
\ee

Now, we construct the following gauge invariant operators,
\be
    \bar{\psi}^{a}_n = \psi^{a}_n \prod_{m>n} V_m.
\ee
We can verify that $[\psi^{a}_n,G_\alpha]=0$ for all $\alpha$. 

Now, we wish to construct a gauge-invariant QCA. We do so by constructing it using gauge invariant operators \textemdash $\bar{\psi}^{a}_n$ and $E_n$.

Let us now define a gauge-invariant unitary $\widetilde{T}$.
\begin{align}
    \widetilde{T} \barPsiB{n} \widetilde{T}^\dagger &= \Pi_a\Pi_b\barPsiB{n+1}+(\Pi_{a}\Pi_{\bar{b}}
    \nonumber\\
    &+\Pi_{\bar{a}}\Pi_{{b}})\barPsiB{n}+\Pi_{\bar{a}}\Pi_{\bar{b}}\barPsiB{n-1}.
    \label{eq:TQCAInvariant}
\end{align}
Rewriting,
\begin{align}
    \widetilde{T} \PsiB{n} \widetilde{T}^\dagger &= \Pi_a\Pi_b\PsiB{n+1}V_{n+1/2}^\dagger+(\Pi_{a}\Pi_{\bar{b}}
    \nonumber\\
    &+\Pi_{\bar{a}}\Pi_{{b}})\PsiB{n}+\Pi_{\bar{a}}\Pi_{\bar{b}}\PsiB{n-1}V_{n-1/2},
    \label{eq:TQCAlocal}
\end{align}
we can see that it is indeed local.

Now, note that $W$, as defined before, is indeed gauge-invariant. We can rewrite $W$ by replacing the unbarred operators in its definition with barred operators to make its gauge invariance explicit. 

Consider $D_{\text{Dirac}}=W\widetilde{T}$. This is a gauge-invariant version of the Dirac QCAs. With a choice of the projection operators, we can obtain a gauge-invariant version of either the conventional Dirac QCA or the more general family of Dirac QCAs. 

Now, so far, we have not introduced interactions. Introducing a kinetic term for the electric field induces interactions;
\be
    D_{E} = \bigotimes_{n\in\Zb+1/2}\Exp{-i\frac{g^2 c^2 \delta t^3 E_n^2}{2 \epsilon_0   \delta x} }
\ee
where $g$ is the charge of the fermionic particle and $\epsilon_0$ the permittivity of free space.

Then, 
\be
    D = D_{\text{Dirac}}D_{E}
\ee
defines a gauge-invariant QCA with electric interactions in 1+1-D. Thus, we have obtained a family of QCAs that are a discrete spacetime version of the Schwinger model. This is a generalisation of the QCA presented in Ref.~\cite{arrighi2020quantum}.

Finally, we must discuss Gauss's law. Consider the following operator
\begin{align}
    J_n &= E_{n+1/2}-E_{n-1/2}-\PsiB{n}^\dagger\PsiB{n}
    \nonumber\\
    &=E_{n+1/2}-E_{n-1/2}-\barPsiB{n}^\dagger\barPsiB{n}.
\end{align}

In the Schr\"odinger picture, we demand
\be
    J_n\ket{\text{phys}} = 0
\ee
for any physical state $\ket{\text{phys}}$. Ensuring that the initial state is physical is sufficient. This is because $J_n$ is the generator of gauge transformations. By demanding that our QCA is gauge invariant, we have already insured that the unitary defining the QCA commutes with $J_n$.

Finally, note that, to implement this QCA on a physical system, we can restrict $l$ to be between $-L$ and $L$ where $L$ is some integer.

\bibliography{myref}
\end{document}